\documentclass[aps,prl,a4paper,twocolumn,floatfix,showpacs,superscriptaddress,preprintnumbers,10pt]{revtex4-2}

\usepackage{amssymb}
\usepackage{graphicx}
\usepackage{xcolor}
\definecolor{gray}{rgb}{0.75, 0.75, 0.75}
\usepackage{upgreek}
\usepackage[utf8]{inputenc}
\usepackage[normalem]{ulem}
\usepackage{hyperref}
\usepackage{siunitx}
\usepackage{layouts}
\usepackage[percent]{overpic}

\usepackage[margin=1in]{geometry}
\usepackage{dcolumn}
\usepackage{bm}
\usepackage{amsmath}

\begin{document}

\title{Active Brownian Dynamics from a Hamiltonian Model: Transitioning from Equilibrium to Activity}

\author{Antik Bhattacharya}
\email{abhattacharya@tifrh.res.in }
\affiliation{Tata Institute of Fundamental Research Hyderabad, 36/P, Gopanpally Village, Serilingampally Mandal, Ranga Reddy District, Hyderabad, Telangana 500046, India}
\author{Smarajit Karmakar}
\email{smarajit@tifrh.res.in}
\affiliation{Tata Institute of Fundamental Research Hyderabad, 36/P, Gopanpally Village, Serilingampally Mandal, Ranga Reddy District, Hyderabad, Telangana 500046, India}
\author{J\"urgen Horbach}
\email{Juergen.Horbach@uni-duesseldorf.de}
\affiliation{Institut f\"ur Theoretische Physik II: Weiche Materie, Heinrich-Heine-Universit\"at D\"usseldorf, Universit\"atsstra\ss e 1, 40225 D\"usseldorf, Germany}

\begin{abstract}
We introduce a Hamiltonian Active Brownian Particle (HABP) model that connects equilibrium dynamics with the non-equilibrium, two-dimensional overdamped behavior of standard Active Brownian Particles (ABPs). In equilibrium, the system follows overdamped Langevin equations that strictly satisfy the fluctuation-dissipation theorem. 
Coupling translational and rotational degrees of freedom to separate heat baths ($T_{\theta} > T_{\textrm{tr}}$), drives the system out of equilibrium. In free space, matching the diffusion coefficients yields quantitative agreement with the ABP model in the limit $T_\theta/T_{\textrm{tr}}\to \infty$, whereas in a harmonic potential, this matching is recovered even at finite temperature ratios.
Entropy production analysis demonstrates that in the ABP limit, all energy injected by the swim force dissipates into the translational bath, leaving the rotational bath as a zero-cost entropy source. These insights provide the first steps toward equilibrium-inspired descriptions of active matter, facilitating the development of new theoretical frameworks to study activity-induced fluctuations, transport, and phase behavior in living and non-living soft matter systems far from equilibrium.

\end{abstract}

\maketitle

{\bf Introduction.} Active systems comprise self-propelled entities that convert locally absorbed energy into directed mechanical work \cite{Vicsek1995, Toner1998, Ramaswamy2010, Marchetti2013, Bechinger2016, Liebchen2018}. Driven inherently far from thermodynamic equilibrium, these systems break detailed balance at the microscopic scale \cite{Fodor2016, Bowick2022}. While this violation also characterizes passive systems under global external fields, active matter is fundamentally distinct: the drive is intrinsic and lacks a canonical equilibrium (Hamiltonian) reference, rendering linear response theory and conventional statistical mechanics inapplicable \cite{Cengio2019, Bowick2022}. Consequently, active matter lacks a well-defined free energy, leaving unique definitions for mechanical state variables like pressure and chemical potential highly contentious \cite{solon2015pressure, solon2018generalized}.

Prominent models such as the Vicsek model \cite{Vicsek1995}, Active Brownian Particles (ABPs) \cite{Romanczuk2012, Bechinger2016}, Active Ornstein-Uhlenbeck Particles (AOUPs) \cite{Szamel2014, Martin2021}, and non-reciprocal Langevin equations \cite{Fruchart2021, Shi2026} successfully reproduce hallmark features of activity, including non-reciprocal interactions, motility-induced phase separation (MIPS) \cite{Fily2012, Buttinoni2013, Solon2015, Digregorio2018}, flocking \cite{Vicsek1995, Toner1998, Cavagna2014}, and swarming \cite{Romanczuk2012, Marchetti2013, Cavagna2017}. However, without an underlying thermodynamic potential, a rigorous classification of these non-equilibrium phenomena remains elusive. The search for equilibrium analogues of active systems is motivated by an even broader question: can the emergent dynamics of active matter originate from an underlying Hamiltonian system with appropriate internal degrees of freedom? This question is significant both conceptually and because it could allow the use of equilibrium statistical mechanics for nonequilibrium systems. Recent studies have identified connections between active dynamics and Hamiltonian formulations. For instance, path probabilities of active Brownian motion can be mapped to the equilibrium statistics of semiflexible polymers, directly linking nonequilibrium trajectories to equilibrium states \cite{shee2020mapping}.

Overcoming the lack of a thermodynamic framework for active systems requires a well-defined equilibrium reference model from which active dynamics can be systematically derived, establishing a direct mathematical bridge to classify active phenomena relative to a passive reference. While recent approaches have demonstrated how various active phenomena can emerge from Hamiltonian descriptions \cite{Casiulis2020, Fieguth2022, Fieguth2024, Bhattacharya2025, Bhattacharya2026, Shi2026, Casiulis2026, Chen2026}, a fundamental mapping between stochastic equations of activity such as the ABP model and an underlying Hamiltonian framework is hitherto lacking.

In this Letter, we provide this crucial connection by introducing a minimal Hamiltonian framework that systematically maps onto the stochastic dynamics of an ABP. We consider a particle at position $\vec{r}$ with momentum $\vec{p}$, coupled to an internal spin degree of freedom described by the unit vector $\vec{S} = (\cos\theta, \sin\theta)^{\textrm{T}}$, where the orientation angle $\theta$ is canonically conjugate to the angular momentum $L$. The total Hamiltonian is defined as
\begin{equation}
\mathcal{H} = \frac{\vec{p}^{\, 2}}{2m} + \frac{L^2}{2I} -  K \vec{S} \cdot \vec{r}, 
\label{eq_hamilton} 
\end{equation} 
where $m$ is the mass, $I$ is the moment of inertia, and $K$ is a coupling constant. Starting from Eq.~\eqref{eq_hamilton}, the overdamped Langevin equations for the coupled translational and rotational degrees of freedom are established by introducing the respective friction forces, $-\gamma_{\textrm{tr}} \dot{\vec{r}}$ and $-\gamma_\theta \dot{\theta}$ (where $\gamma_{\textrm{tr}}$ and $\gamma_{\theta}$ are the translational and rotational friction coefficients), along with uncorrelated Gaussian white noise sources, and subsequently neglecting inertial terms. 

Crucially, we generalize this framework by coupling the spatial and angular dynamics to distinct thermal baths at temperatures $T_{\textrm{tr}}$ and $T_{\theta}$, respectively \cite{Netz2020}. While equilibrium is preserved under isothermal conditions ($T_{\textrm{tr}} = T_\theta$), a non-isothermal setting ($T_{\textrm{tr}} \neq T_\theta$) combined with an appropriate choice of friction coefficients exactly maps the system's effective dynamics onto those of an ABP within a well-defined and controllable time regime. Applying this mapping to both free and harmonically trapped particles, we uncover the microscopic origin of the emergent active drive: it stems from an inter-bath heat flux, mediated by the potential energy term $- K \vec{S} \cdot \vec{r}$ acting as an energetic transducer between the degrees of freedom. We quantify the inherent irreversibility of this non-isothermal transport by deriving an explicit expression for the steady-state entropy production.

{\bf Equations of motion.} 
The stochastic equations of motion, derived from the Hamiltonian (\ref{eq_hamilton}) and generalized to couplings with two heat baths at temperatures $T_{\textrm{tr}}$ and $T_{\theta }$, are given by
\begin{eqnarray}
\dot{\vec{r}} & = & 
\frac{K}{\gamma_{\textrm{tr}}} \vec{S} + \sqrt{\frac{2k_BT_{\textrm{tr}}}{\gamma_{\textrm{tr}}}} \; 
\vec{\xi}(t) , \label{eq_dr} \\
\dot{\theta} & = & 
\frac{K}{\gamma_{\theta}} \vec{S}^\perp \cdot \vec{r} + 
\sqrt{\frac{2k_BT_{\theta}}{\gamma_{\theta}}} \; \eta(t), 
\label{eq_dtheta}
\end{eqnarray}
where $\vec{S}^\perp = \frac{\partial \vec{S}}{\partial \theta}$; $\vec{\xi}(t)$ and $\eta(t)$ are independent Gaussian white noise \cite{risken1989fokker} sources satisfying
\begin{eqnarray}
\langle \xi_\alpha(t) \, \xi_\beta(t') \rangle & = &
\delta_{\alpha\beta} \delta(t-t'), \label{eq_ntheta}\\ 
\langle \eta(t) \, \eta(t') \rangle & = & \delta(t-t'). \label{eq_ntrans}
\end{eqnarray}
For $T_{\textrm{tr}}=T_\theta$, the noise amplitudes in Eqs.~(\ref{eq_dr}) and (\ref{eq_dtheta}) ensure that the fluctuation-dissipation theorem (FDT) is satisfied, guaranteeing that detailed balance holds. Conversely, for $T_{\textrm{tr}}\neq T_\theta$, a net heat flux arises between the two thermal baths, mediated by the torque $K \vec{S}^\perp \cdot \vec{r}$ in Eq.~\eqref{eq_dtheta}. This breaks detailed balance and drives the system into non-equilibrium states. In the following, the model defined by Eqs.~(\ref{eq_dr})-(\ref{eq_ntrans}) is referred to as the Hamiltonian Active Brownian Particle (HABP) model.

Indeed, the HABP model shares similarities with the standard ABP model for which the stochastic equations of motion read
\begin{eqnarray}
\dot{\vec{r}} & = & 
v_0 \vec{S} + \sqrt{2D_{\textrm{tr}}} \; 
\vec{\xi}(t) , \label{eq_abp_dr} \\
\dot{\theta} & = & 
\sqrt{2 D_\theta} \; \eta(t), 
\label{eq_abp_dtheta}
\end{eqnarray}
where $v_0$ is the propulsion speed, while $D_{\textrm{tr}}$ and $D_\theta$ denote the translational and rotational diffusion coefficients, respectively \cite{Romanczuk2012, Bechinger2016}. The HABP equations \eqref{eq_dr} and \eqref{eq_dtheta} qualitatively differ from the ABP dynamics due to the additional deterministic torque term $\propto \vec{S}^\perp \cdot \vec{r}$ in Eq.~\eqref{eq_dtheta}, which couples orientation to position: the spin reorients according to the particles displacement with respect to the origin. In contrast, the angular velocity $\dot{\theta}$ in the standard ABP model is solely driven by white noise, implying that the orientation performs free rotational Brownian motion entirely decoupled from position. Unlike Eqs.~(\ref{eq_dr}) and (\ref{eq_dtheta}) at thermal equilibrium ($T_{\textrm{tr}}=T_\theta$), the stationary state of an ABP never satisfies detailed balance \cite{Fodor2016, Shankar2018} for any non-zero choice of the parameters $v_0$, $D_{\textrm{tr}}$ and $D_\theta$. In turn, we note that no Hamiltonian exists that simultaneously produces a propulsive force $v_0\vec{S}$ and zero torque on $\theta$. This would violate the integrability condition $\frac{\partial}{\partial \theta} \left( \frac{\partial \mathcal{H}}{\partial r_\alpha} \right) =  \frac{\partial}{\partial r_\alpha} \left( \frac{\partial \mathcal{H}}{\partial \theta} \right)$.

\begin{figure}
\vskip +0.2in
\centering
\includegraphics[width=0.49\linewidth]{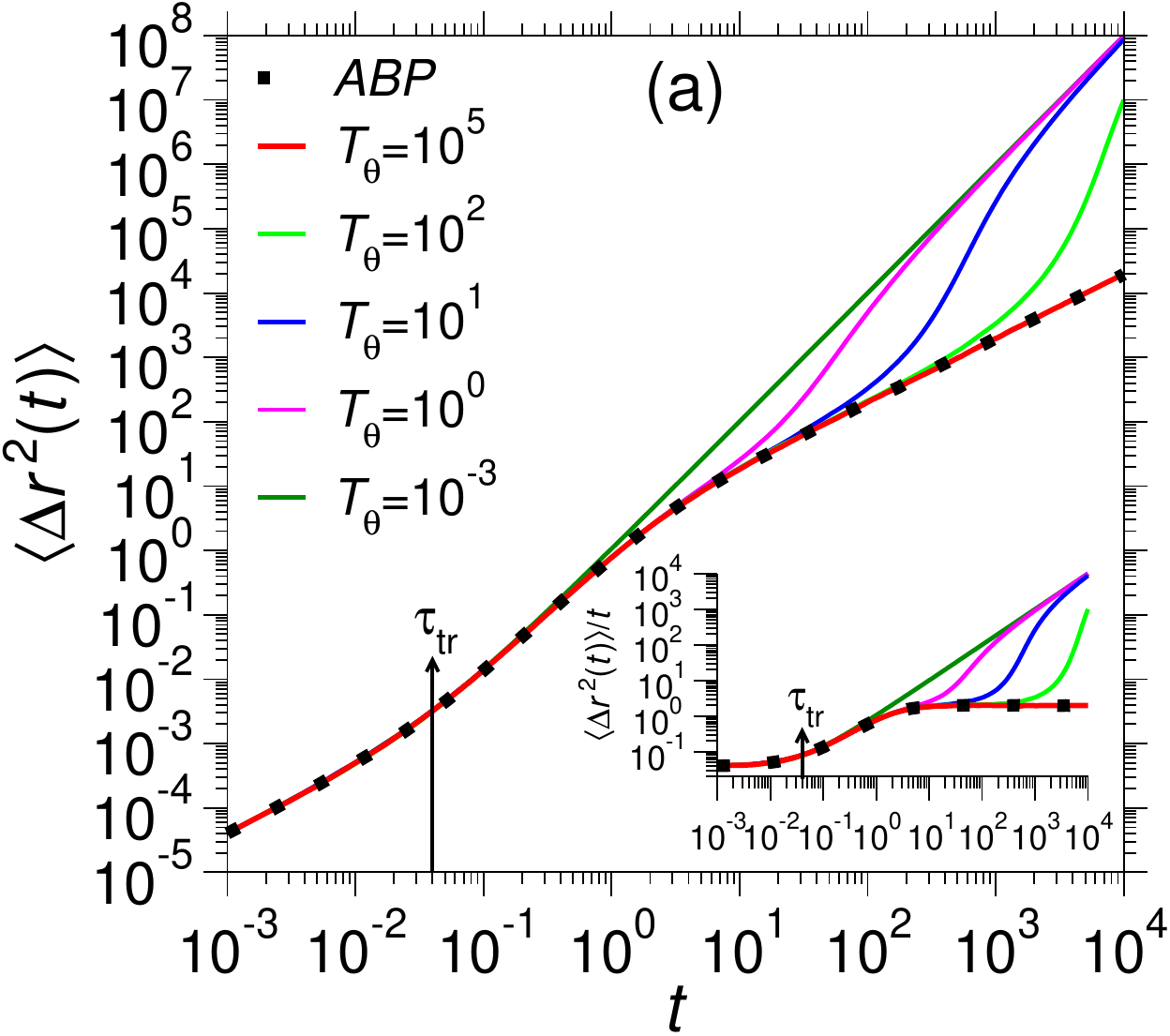}
\includegraphics[width=0.49\linewidth]{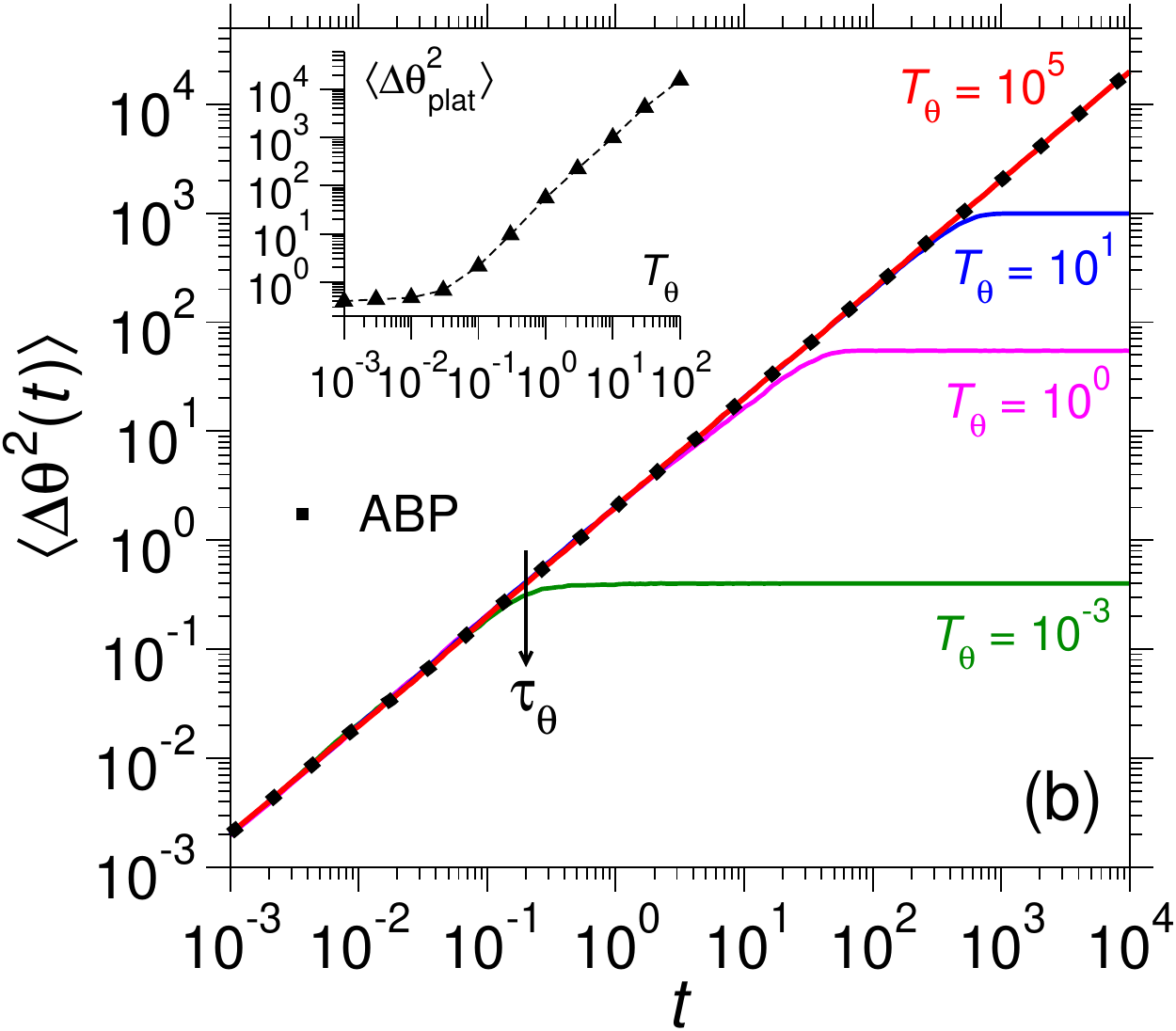}
\caption{(a) Translational MSD, $\langle \Delta \vec{r}^{\, 2}(t) \rangle$, for the HABP model for $D_{\textrm{tr}}=0.01$ (with $T_{\textrm{tr}}=10^{-3}$ fixed) and $D_\theta = 1.0$, varying $T_\theta$ from $10^{-3}$ to $10^{5}$ (see text). The black dotted line is the corresponding ABP result for $D_{\textrm{tr}}=0.01$, $D_\theta = 1.0$, and $v_0 = 1.0$. The inset shows the same data but now $\langle \Delta \vec{r}^{\, 2}(t) \rangle/t$ is plotted as function of time. (b) Same as a), but now for the rotational MSD, $\langle \Delta \theta^2(t) \rangle$. The inset shows the plateau height $\langle \Delta \theta^2_{\textrm{plat}}\rangle$ as a function of $T_\theta$. \label{fig1}}
\end{figure}
{\bf Free-particle simulation of ABP and HABP and choice of parameters.} 
As a next step, we compare the ABP and HABP models via computer simulations, analyzing their distinct time regimes. To this end, both Eqs.~\eqref{eq_dr} and \eqref{eq_dtheta} for the HABP and Eqs.~\eqref{eq_abp_dr} and \eqref{eq_abp_dtheta} for the ABP were solved numerically using an Euler–Maruyama algorithm \cite{kloeden1977numerical}. Comparing the respective equations allows us to identify the translational and rotational diffusion coefficients of the ABP model as $D_{\textrm{tr}}=k_{\textrm{B}} T_{\textrm{tr}}/\gamma_{\textrm{tr}}$ and $D_\theta= k_{\textrm{B}} T_\theta/\gamma_\theta$, respectively, while the propulsion speed is given by $v_0=K/\gamma_{\textrm{tr}}$. In the following, we set the coupling constant to $K=0.1$. The diffusion coefficients are kept fixed at $D_{\textrm{tr}}= 0.01$ and $D_{\theta}=1.0$, while the rotational temperature $T_{\theta }$ is varied relative to the fixed translational temperature $T_{\textrm{tr}}=10^{-3}$. Consequently, for $T_\theta = T_{\textrm{tr}} = 10^{-3}$, the friction coefficients are $\gamma_{\textrm{tr}}=0.1$ and $\gamma_\theta = 10^{-3}$, whereas for $T_\theta = 10^5$, e.g., we set $\gamma_\theta = 10^5$. As shown below, this parameter selection enables a controlled interpolation between equilibrium and the ABP-like non-equilibrium dynamics in the framework of the HABP model.
 
{\bf Single free particle.} To establish the precise connection between the ABP and the HABP out of equilibrium ($T_{\textrm{tr}}\neq T_\theta$), we first analyze the translational and rotational mean-squared displacement (MSD) of a single free particle. These are defined as $\langle \Delta \vec{r}^{\, 2}(t)\rangle = \langle (\vec{r}(t) - \vec{r}_0)^2 \rangle$ and $\langle \Delta \theta^2(t) \rangle = \langle (\theta(t) - \theta(0) )^2 \rangle$, respectively, where $\langle \dots \rangle$ denotes an average over a large number of independent particle trajectories (typically $10^7$) and $\vec{r}_0 = (0, 0)^{\textrm{T}}$ is the origin from which the particle starts at time $t=0$. Figure \ref{fig1} displays both MSDs for the HABP model across a range of rotational temperatures, $10^{-3} \le T_\theta \le 10^5$, with all other parameters fixed as described previously. The corresponding results for the standard ABP model are included as dotted lines in Figs.~\ref{fig1}(a) and \ref{fig1}(b).

At thermal equilibrium ($T=T_{\textrm{tr}} = T_\theta$), the translational MSD for the HABP model exhibits a crossover from a short-time diffusive behavior ($\propto t$) to a ballistic regime ($\propto t^2$) in the long-time limit. Concurrently, the rotational MSD transitions from a linear diffusive dependence on $t$ at short times to a constant plateau, $\langle \Delta \theta_{\textrm{plat}}^2 \rangle$, at long times. This saturation indicates that the torque $\propto \vec{S}^\perp \cdot \vec{r}$ effectively locks the spin orientation and thus the angle $\theta$ in the long-time limit. This locking directly dictates translational dynamics: at long times, the fixed orientation induces a ballistic behavior $\sim \frac{K^2}{\gamma_{\textrm{tr}}^2} t^2$, whereas the short-time dynamics remain purely diffusive with $\langle \Delta \vec{r}^{\,2}(t) \rangle= \frac{4 k_{\textrm{B}}T}{\gamma_{\textrm{tr}}} t$. Equating the short-time and long-time scaling behaviors yields the characteristic crossover timescale $\tau_{\textrm{tr}} = \frac{4 k_{\textrm{B}}T \gamma_{\textrm{tr}}}{K^2}$, which is explicitly indicated in Fig.~\ref{fig1}(a). Similarly, the crossover timescale in the rotational MSD, $\tau_\theta$, can be estimated via the time where short time diffusive regime reaches the plateau value $\langle \Delta \theta_{\textrm{plat}}^2 \rangle$ and thus $\tau_\theta = \frac{\gamma_\theta \langle \Delta \theta_{\textrm{plat}}^2 \rangle}{2 k_{\textrm{B}}T}$ (this timescale is indicated in Fig.~\ref{fig1}(b)). The value of the plateau is approximately given by $\langle \Delta \theta_{\textrm{plat}}^2 \rangle \approx \frac{4 k_{\textrm{B}}T}{K} \sqrt{\frac{\gamma_{\textrm{tr}}}{\gamma_\theta}}$ and thus $\tau_\theta \approx \frac{2\gamma_\theta}{K} \sqrt{\frac{\gamma_{\textrm{tr}}}{\gamma_\theta}}$ (see supplemental material).

\begin{figure}
\centering
\includegraphics[width=0.49\linewidth]{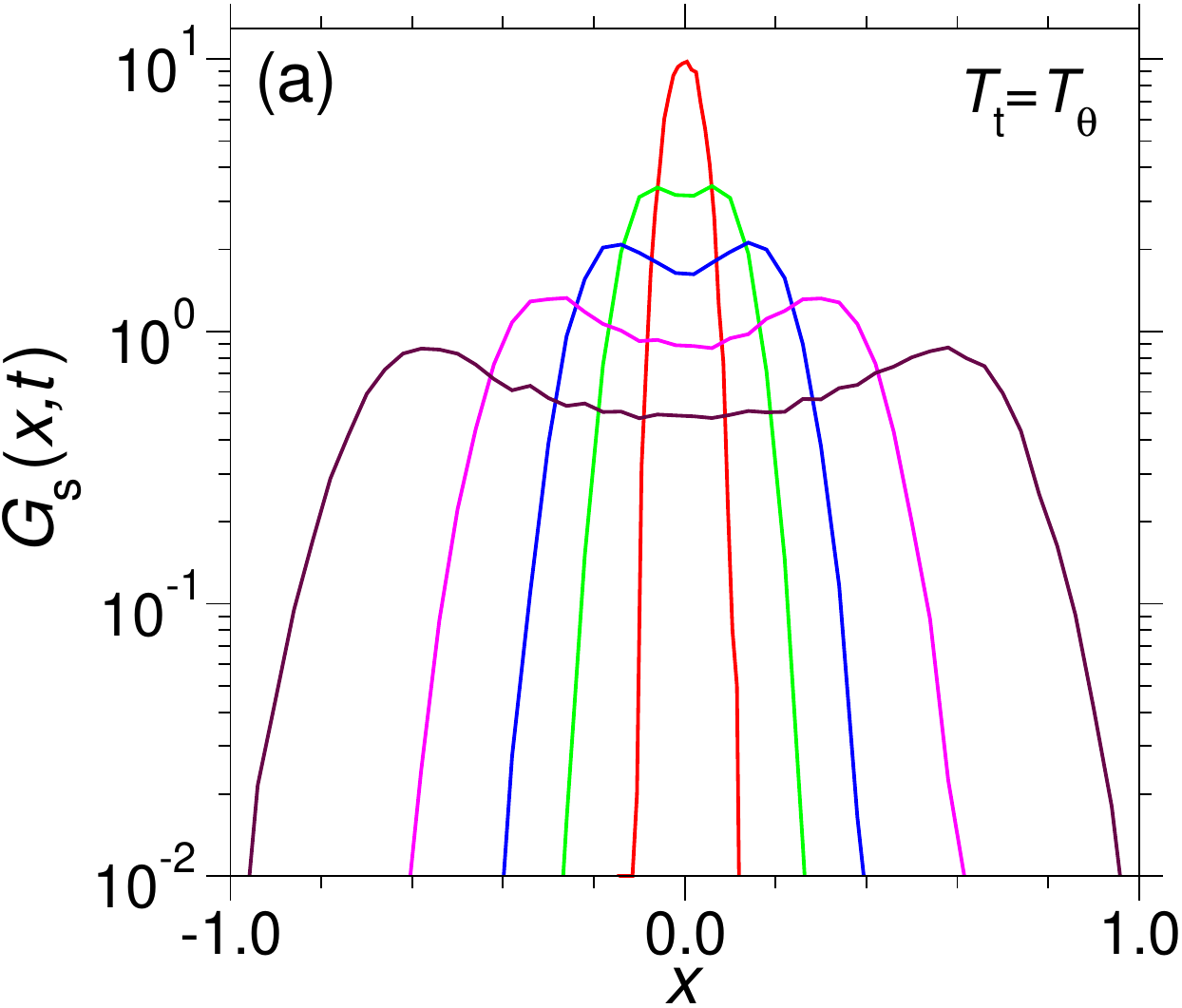}
\includegraphics[width=0.49\linewidth]{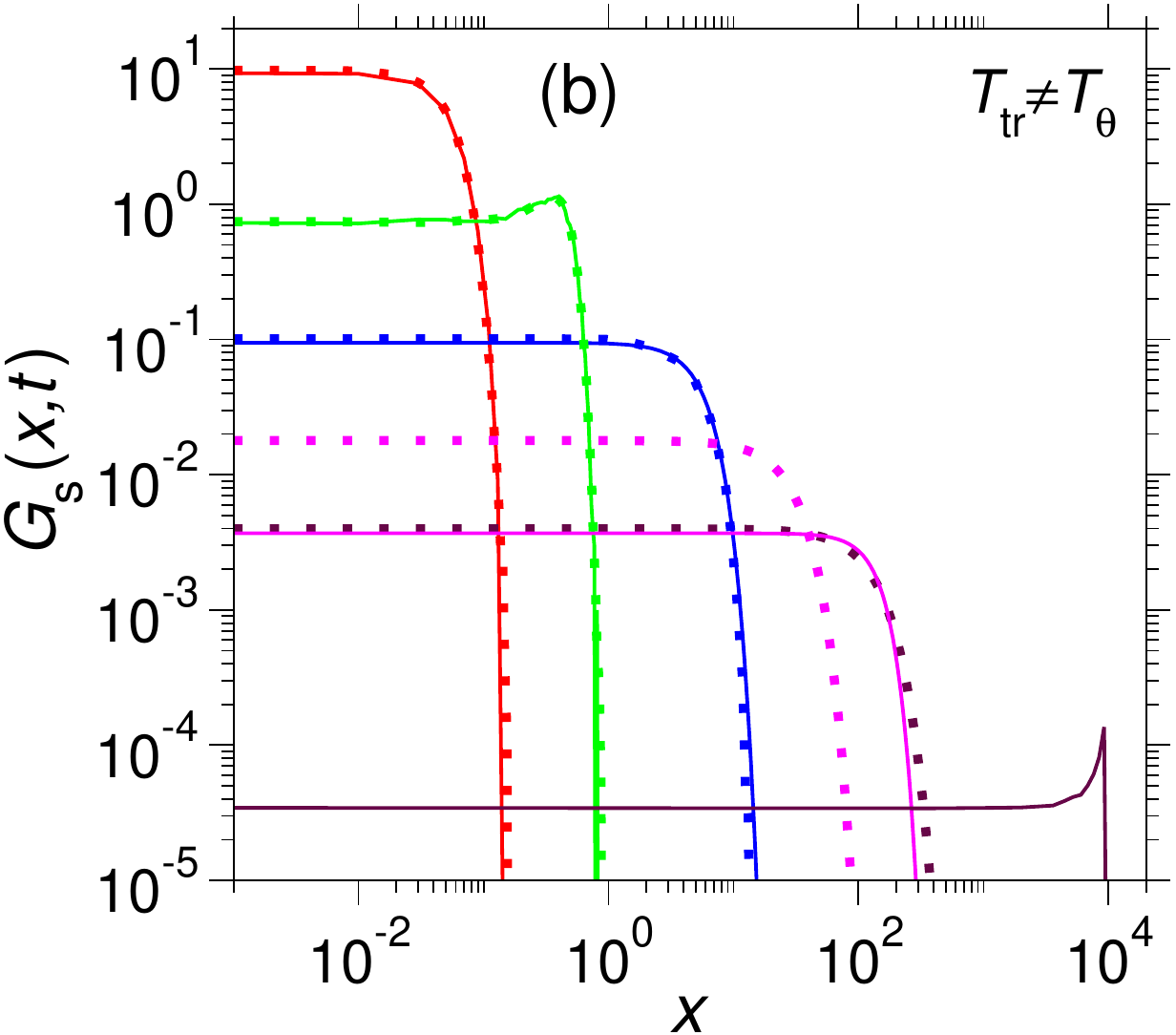}
\caption{(a) Van Hove correlation, $G_s(x, t)$, for $T_{\textrm{tr}}=T_\theta= 10^{-3}$ (equilibrium)  and the times $t=0.036$, $0.115$, $0.205$, $0.365$, and $0.652$ (using the color gradient from red to brown with increasing time). (b) $G_s(x, t)$ for $T_\textrm{tr}=10^{-3}$ and $T_\theta = 10^2$ at times $t=0.038$, $0.483$, $15.0$, $483.33$, and $10000.0$ (keeping $D_{\textrm{tr}}=0.01$ and $D_\theta = 1.0)$. The dotted lines are the corresponding ABP results, in quantitative agreement with HABP up to $t=15.0$. \label{fig2}}
\end{figure}
For the ABP model, the mean squared displacements (MSDs) of a free particle are analytically given by $\langle \Delta \theta^2(t) \rangle = 2 D_\theta t$ and $\langle \Delta \vec{r}^{\, 2}(t) \rangle= 4 D_{\textrm{tr}} t + \frac{2 v_0^2}{D_{\theta}^2} (D_\theta t - 1 + {\rm e}^{-D_\theta t})$ \cite{Bechinger2016, Howse2007, Romanczuk2012}. For the translational MSD, this implies a ballistic regime at intermediate timescales ($\tau_{\textrm{tr}} < t < 1/D_\theta$), where $\langle \Delta \vec{r}^{\, 2}(t) \rangle \approx v_0^2 t^2$. In the long-time limit, a diffusive crossover is recovered, yielding $\langle \Delta \vec{r}^{\, 2}(t) \rangle = 4 D_{\textrm{eff}} t$ with the effective diffusion coefficient $D_{\textrm{eff}} = D_{\textrm{tr}} + \frac{v_0^2}{2 D_\theta}$. The physical mechanism underlying this intermediate ballistic regime is straightforward: here, the particle's orientation is essentially frozen because rotational diffusion has not yet had sufficient time to reorient it. Consequently, the particle moves persistently along its instantaneous heading direction with velocity $v_{0}$.

Replicating this behavior in the HABP system (namely, a short-time diffusive behavior followed by an intermediate ballistic regime and a late-stage diffusive regime with coefficient $D_{\textrm{eff}}$ requires delaying the orientational locking caused by the torque $\propto \vec{S}^\perp \cdot \vec{r}$. As demonstrated below, this is only achievable under non-equilibrium conditions ($T_{\textrm{tr}} \neq T_\theta$) by choosing $T_\theta \gg T_{\textrm{tr}}$ while
keeping $T_{\textrm{tr}}$, $D_{\textrm{tr}}$, and $D_{\theta}$ fixed at the values defined above.

These dynamics are illustrated in Fig.~\ref{fig1}. As $T_{\theta}$ increases, the characteristic timescale $\tau_{\theta}$ grows, and the long-time rotational plateau $\langle \Delta \theta^2_{\textrm{plat}} \rangle$ shifts toward larger values. As shown in the inset of Fig.~\ref{fig1}(b), $\langle \Delta \theta^2_{\textrm{plat}}\rangle$ exhibits a rapid increase as a function of $T_{\theta}$ for $T_\theta \gtrsim 10^{-2}$. Because $\tau_\theta \sim \langle \Delta \theta^2_{\textrm{plat}} \rangle$, the crossover timescale where the translational MSD enters the asymptotic ballistic regime shifts significantly to later times with increasing $T_{\theta}$ (Fig.~\ref{fig1}(a)). Therefore, when holding $D_\theta = k_{\textrm{B}} T_\theta/\gamma_\theta$ constant, the temperature $T_{\theta}$ acts as the control parameter that dictates the timescale up to which the HABP dynamics quantitatively match the standard ABP model.

The translational MSD in $x$ direction corresponds to the second moment of the self part of the van Hove correlation function, $G_{\textrm{s}}(x,t) = \langle \delta(x-\Delta x(t))\rangle$\cite{hansen2013theory}, which gives the probability density for a displacement $\Delta x(t) = x(t) - x(0)$ along the $x$-axis at time $t$. In Fig.~\ref{fig2}(a), $G_{\textrm{s}}(x,t)$ is plotted for the equilibrium state ($T_{\textrm{tr}} = T_\theta = 10^{-3}$) at the indicated times. For $t < \tau_\theta = 0.4$, the distribution is Gaussian, characterizing short-time diffusion. For $t > \tau_\theta$, however, it becomes bimodal due to persistent ballistic motion with equal probability in the $+x$ and $-x$ directions. The non-equilibrium case ($T_\theta = 10^2$) in Fig.~\ref{fig2}(b) reveals quantitative agreement between the HABP and ABP models for all evaluated times ($0.036 \le t \le 0.652$), matching the behavior of the translational MSD seen in Fig.~\ref{fig1}(a).

\begin{figure}
\centering
\includegraphics[width=0.49\linewidth]{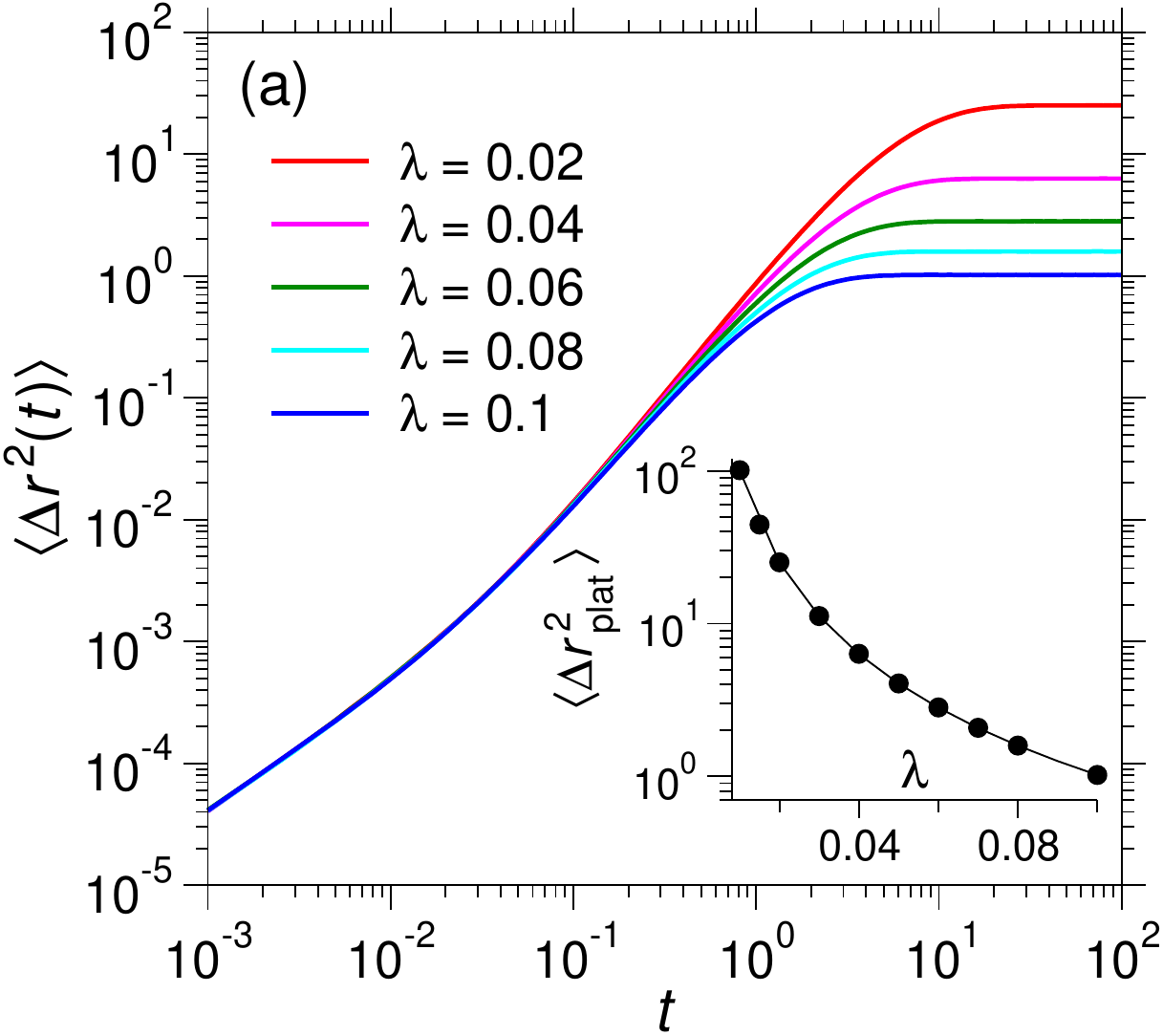}
\includegraphics[width=0.49\linewidth]{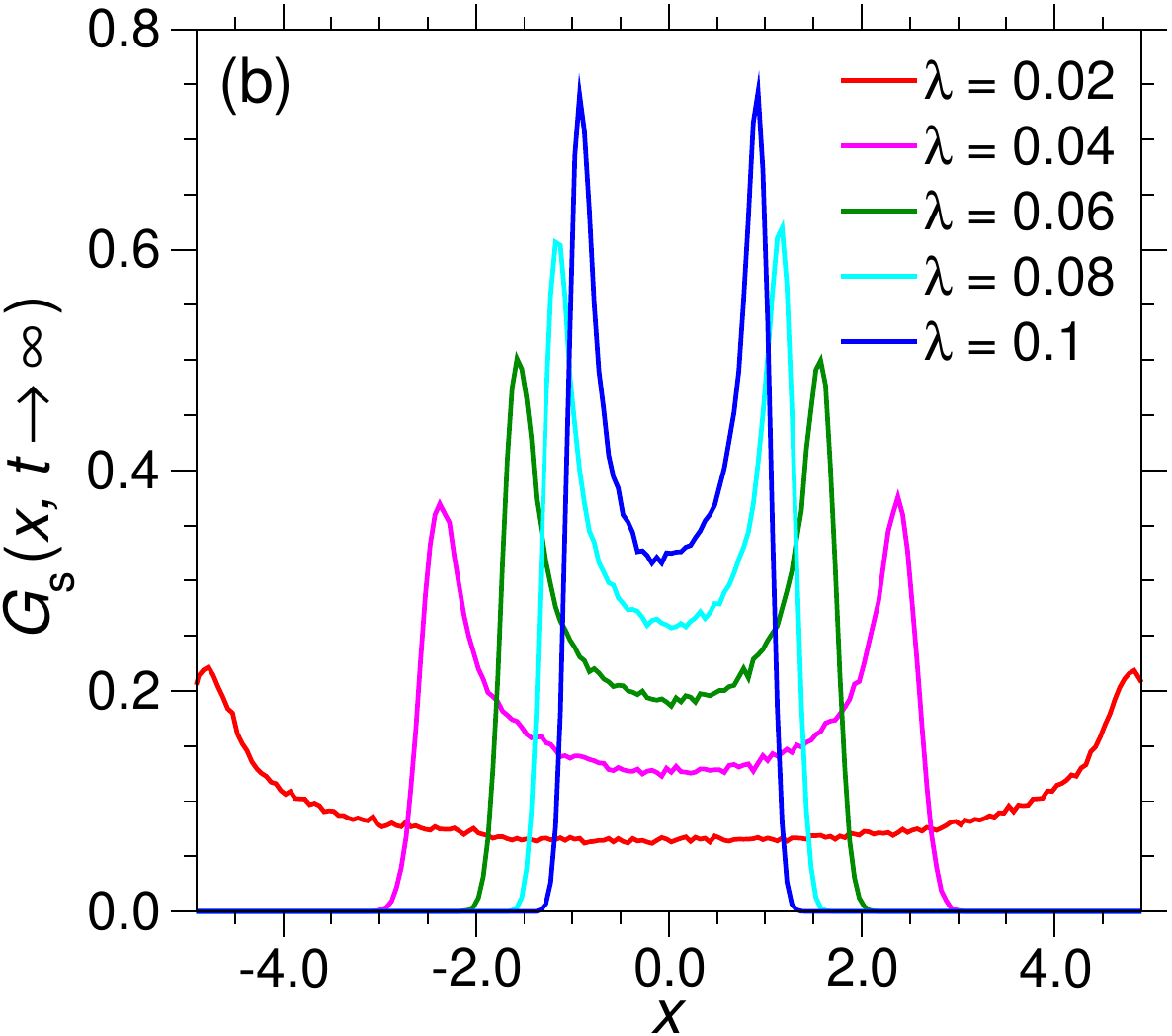}
\includegraphics[width=0.49\linewidth]{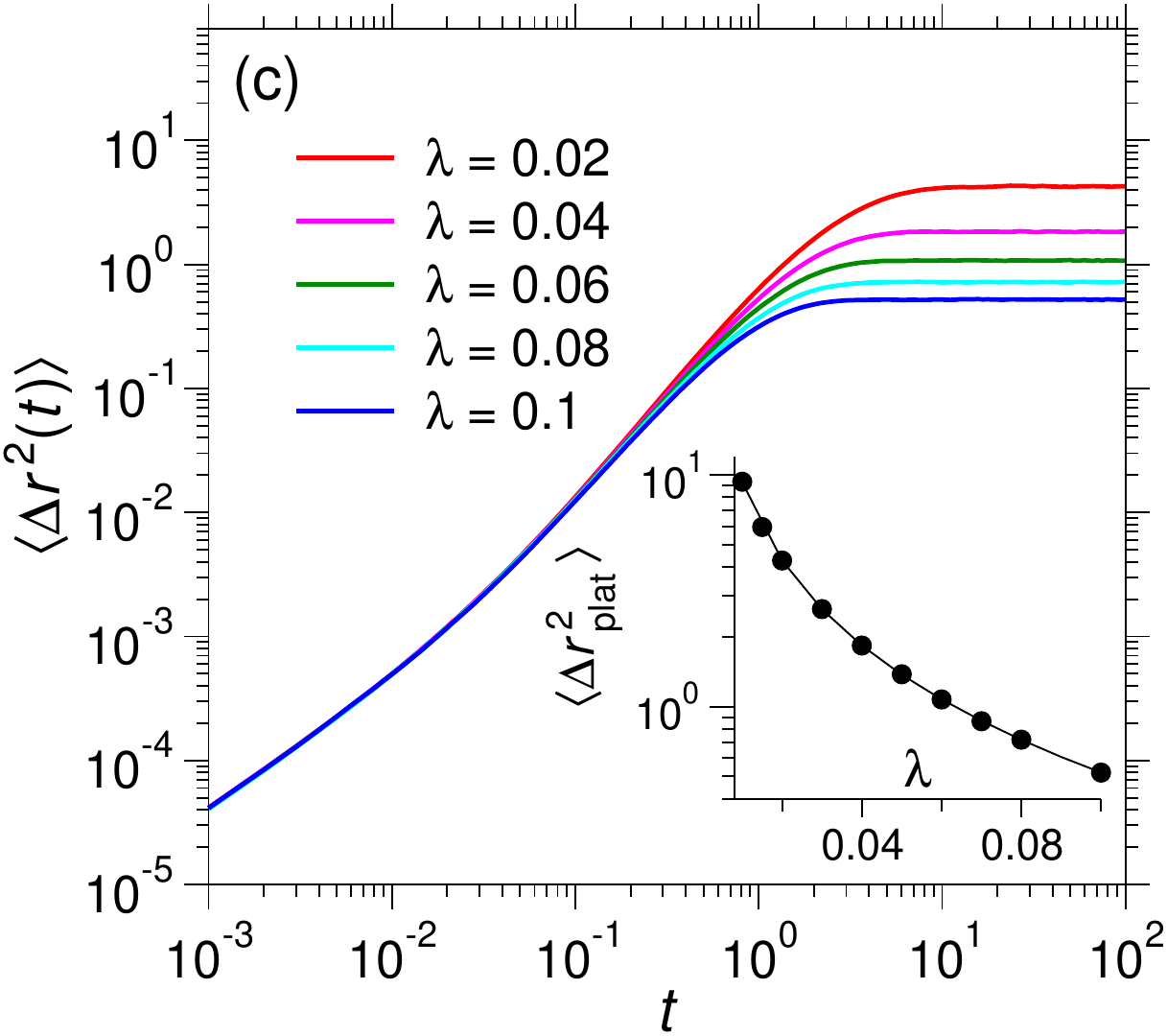}
\includegraphics[width=0.49\linewidth]{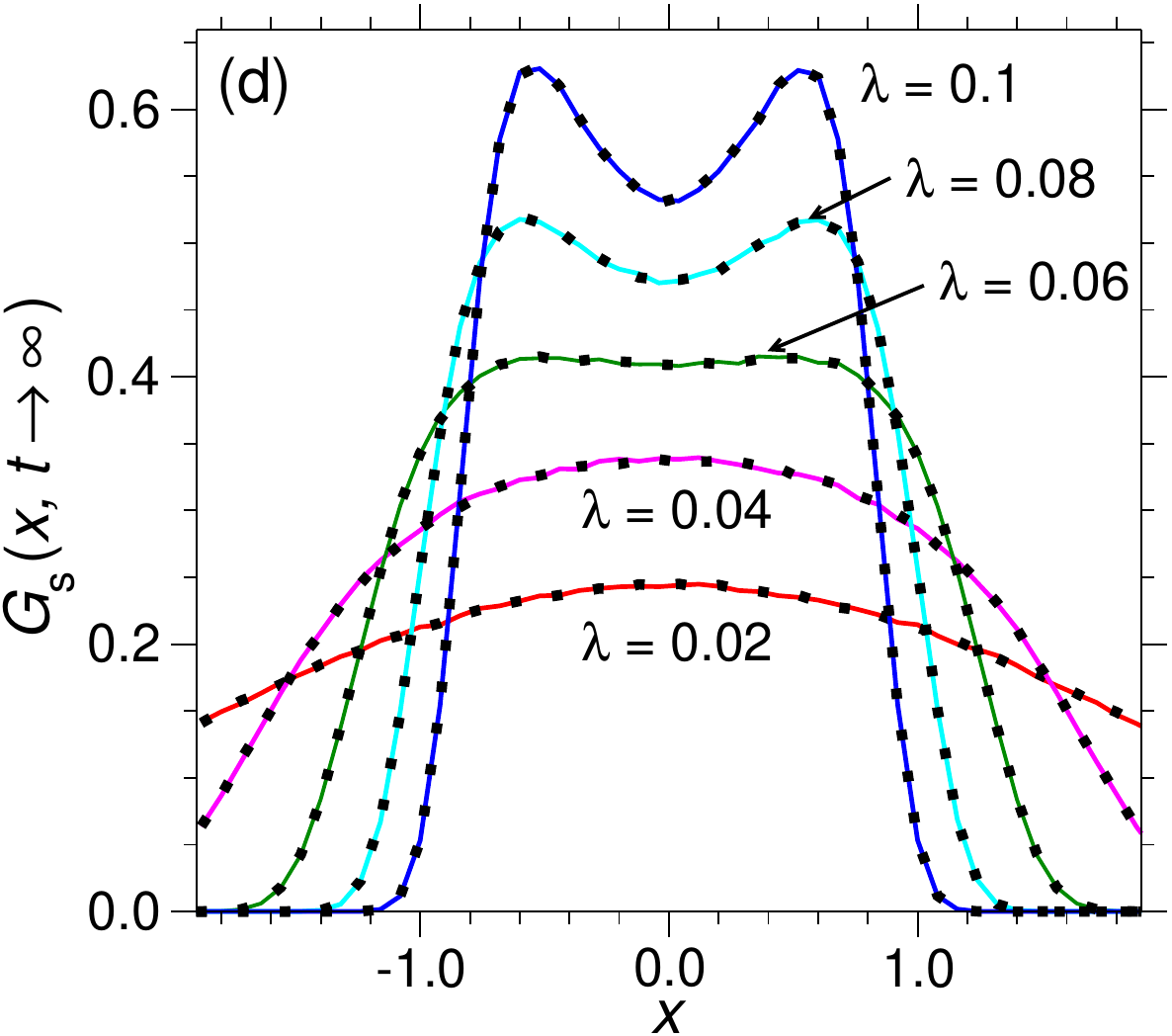}
\caption{ {\it HABP/ABP in a harmonic potential.} (a) MSD, $\langle \Delta \vec{r}^{\,2}(t) \rangle$, for the equilibrium HABP ($T_\theta = T_{\textrm{tr}} = 10^{-3}$) at various trap strengths $ 0.02 \le \lambda \le 0.1$.  Inset: Plateau value $\langle \Delta r^{\, 2}_{\textrm{plat}} \rangle$ versus $\lambda$; the solid line shows the analytical prediction. (b) Stationary van Hove correlation function $G_{\textrm{s}}(x, t \to \infty)$ for the states in (a). (c) MSD for the non-equilibrium case ($T_\theta = 10^{5} > T_{\textrm{tr}} = 10^{-3}$) across the same range of $\lambda$. Inset: $\langle \Delta r^{\, 2}_{\textrm{plat}} \rangle$ versus $\lambda$ (filled circles) compared to the analytical prediction (solid line; see text). (d) Corresponding stationary van Hove correlation function $G_{\textrm{s}}(x, t \to \infty)$ for the states in (c). Dotted black lines indicate the ABP model. \label{fig3}}
\end{figure}
{\bf Particle in a harmonic potential.} Next, we examine the HABP and ABP models for a particle confined in a harmonic potential by adding the restoring force $-\lambda \vec{r}$ (with $\lambda$ a spring constant or trap strength) to Eqs.~\eqref{eq_dr} and \eqref{eq_abp_dr}, respectively \cite{Pototsky2012, Malakar2020, Chaudhuri2021, Caraglio2022, Nakul2023}. Because the harmonic potential restricts the particle's spatial boundaries, the orientational locking that typically occurs in the HABP system when $T_\theta > T_{\textrm{tr}}$ can be completely suppressed, even at finite temperatures $T_{\theta}$.

Figure~\ref{fig3}(a) shows the translational MSD of the HABP in equilibrium ($T_\theta = T_{\textrm{tr}} = 10^{-3}$) for various trap strengths $\lambda \in [0.02, 0.1]$. Due to harmonic confinement, the MSDs exhibit a plateau at long times. The corresponding stationary van Hove correlation functions, $G_{\textrm{s}}(x,t \to \infty)$ in Fig.~\ref{fig3}(b), are bimodal, reflecting directed particle motion toward the potential boundaries. As derived in Appendix A of the End Matter, the plateau height is given by $\langle \Delta r^2_{\textrm{plat}} \rangle= \frac{K^2}{\lambda^2} + \frac{2 k_{\textrm{B}}T}{\lambda}$ [see inset of Fig.~\ref{fig3}(a)]. Similar non-equilibrium behavior ($T_\theta = 10^5 > T_{\textrm{tr}}$) is observed in Figs.~\ref{fig3}(c) and (d), where the HABP van Hove functions and plateau values quantitatively match the ABP model. As shown in the inset of Fig.~\ref{fig3}(c), our simulation data perfectly agree with the analytical ABP prediction, $\langle \Delta r^{\, 2}_{\textrm{plat}} \rangle = \frac{K^2}{\lambda (\lambda + \gamma_{\textrm{tr}}D_\theta)} + \frac{2 k_{\textrm{B}}T_{\textrm{tr}}}{\lambda}$, as extracted from Refs.~\cite{Chaudhuri2021, Caraglio2022}.  Note that this expression differs from the ones reported in Refs.~\cite{Chaudhuri2021, Caraglio2022} by a factor of 2; this discrepancy arises because our MSD is defined relative to a fixed initial position at the origin, $\vec{r}(0) = \vec{r}_{0} = \vec{0}$, whereas Refs.~\cite{Chaudhuri2021, Caraglio2022} average over a stationary distribution of initial positions.

\begin{figure}
\centering
\includegraphics[width=0.99\linewidth]{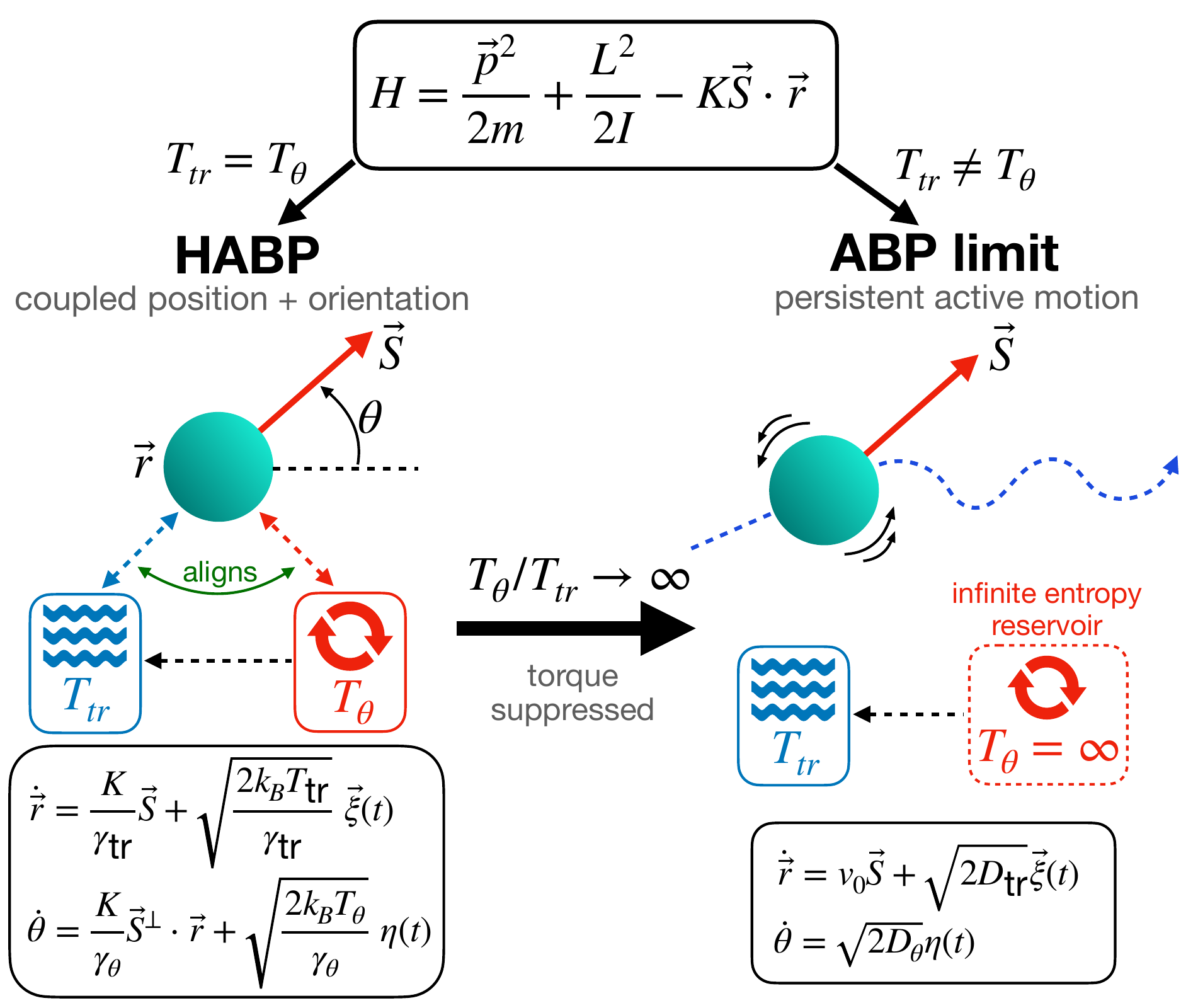}
\caption{Schematic illustration of the transition from equilibrium to active dynamics in the HABP model. \label{fig4}}
\end{figure}

{\bf Entropy production.} For $T_\theta > T_{\textrm{tr}}$, detailed balance in the HABP equations is broken, yielding irreversible stationary states with a non-zero entropy production rate, $\dot{\sigma}$. The entropy production is given by [Appendix B of the End Matter] $\dot{\sigma} = \frac{1}{k_{\textrm{B}}} \left(\frac{1}{T_\theta} - \frac{1}{T_{\textrm{tr}}} \right) \langle \dot{Q} \rangle$, where $\langle \dot{Q} \rangle$ is the mean heat rate mediated by the coupling $K\vec{S}\cdot\vec{r}$.  This heat rate describes the rate of energy flow from the translational to the rotational bath \cite{Sekimoto1998, Seifert2012, Filliger2007}. For a free HABP, it is given by $\langle \dot{Q} \rangle = - K \langle \vec{S} \cdot \dot{\vec{r}} \rangle = K \langle \dot{\vec{S}} \cdot \vec{r} \rangle = - \frac{K^2}{\gamma_{\textrm{tr}}}$, whereas for a harmonically trapped HABP, $\langle \dot{Q} \rangle = \frac{K}{\gamma_{\textrm{tr}}} (K + \lambda \langle \vec{S} \cdot \vec{r} \rangle)$ [for the derivation of these results, see Appendix B of the End Matter]. The resulting entropy production is $\dot{\sigma} = \frac{K^2}{\gamma_{\textrm{tr}} k_{\textrm{B}}} \left( \frac{1}{T_{\textrm{tr}}}- \frac{1}{T_{\theta}} \right)$ for the free particle and $\dot{\sigma} = \frac{K}{\gamma_{\textrm{tr}} k_{\textrm{B}}} (K + \lambda \langle \vec{S} \cdot \vec{r} \rangle) \left( \frac{1}{T_{\textrm{tr}}}- \frac{1}{T_{\theta}} \right)$ for the trapped particle. For $T_\theta \gg T_{\textrm{tr}}$, the stationary correlation simplifies to $\langle \vec{S} \cdot \vec{r} \rangle = - \frac{K}{\lambda + \gamma_{\textrm{tr}} D_\theta}$. Consequently, in the ABP limit ($T_\theta \to \infty$), we recover the established results \cite{Speck2016, Shankar2018, Pietzonka2018, Mandal2017}, applying to the case of time reversible symmetry (TRS)-even propulsion \cite{Shankar2018}: $\dot{\sigma} = \frac{K^2}{\gamma_{\textrm{tr}} k_{\textrm{B}}T_{\textrm{tr}}} = \frac{\gamma_{\textrm{tr}}v_0^2}{k_{\textrm{B}}T_{\textrm{tr}}}$ for the free particle and $\dot{\sigma} = \frac{K^2 D_\theta}{k_{\textrm{B}}T_{\textrm{tr}} (\lambda + \gamma_{\textrm{tr}} D_\theta)}$ for the harmonically trapped particle. In this limit, the rotational bath acts as an infinite entropy reservoir that injects orientational randomness at zero thermodynamic cost (due to its infinite temperature) without changing its own entropy. Furthermore, the harmonic confinement suppresses entropy production, as expected. Note that the ABP limit of the HABP naturally recovers an ABP model with TRS-even propulsion; this arises because the self-propulsion enters via the force $K \vec{S}$, which remains invariant under time reversal.

{\bf Conclusions.} Figure \ref{fig4} summarizes the transition from the equilibrium HABP to the ABP limit ($T_\theta/T_{\textrm{tr}} \to \infty$) while keeping $D_\theta = \frac{k_{\textrm{B}}T_\theta}{\gamma_\theta}$ constant. To achieve non-equilibrium steady states characterized by broken detailed balance and entropy production, the HABP, defined by the Hamiltonian \eqref{eq_hamilton}, is coupled to two heat baths at different temperatures. Activity thus emerges naturally from the thermodynamic asymmetry between the translational and rotational sectors. Consequently, the HABP model replaces the phenomenological swim force of standard active matter with an explicit thermodynamic mechanism. The temperature difference between the baths serves as a tunable parameter that interpolates between equilibrium and active non-equilibrium dynamics. This controlled interpolation enables perturbative and thermodynamic approaches to active matter, which are currently being pursued. 

A particularly important direction for future work is to extend the HABP framework to interacting many-particle systems and investigate whether it can provide a microscopic basis for understanding collective phenomena in dense active matter. Such an extension would allow the systematic study of activity-induced changes in structural correlations, collective fluctuations, and mechanical stability as highlighted recently in \cite{DeyNatComm2025, SharmaNatPhys2025, Bhattacharya2026, RashmiNatComm2026}. In particular, the Hamiltonian formulation may provide a natural route to investigate how active forces couple to long-wavelength density fluctuations, which has been proposed as a key mechanism underlying anomalous fluctuations and the stability of active solids \cite{DeyNatComm2025}. It would also be valuable to explore the relationship between activity and externally driven systems, particularly the proposed correspondence between active forcing and oscillatory shear \cite{SharmaNatPhys2025, GoswamiNatPhys2025, RashmiNatComm2026}. Finally, these developments could establish Hamiltonian active models as a unifying reference framework for connecting microscopic activity, collective fluctuations, yielding, and mechanical response across a broad class of active materials.

{\bf Acknowledgments.} We acknowledge funding by intramural funds at TIFR Hyderabad from the Department of Atomic Energy (DAE) under Project Identification No. RTI 4007. SK acknowledges the Swarna Jayanti Fellowship grants DST/SJF/PSA01/2018-19 and SB/SFJ/2019-20/05 from the Science and Engineering Research Board (SERB) and Department of Science and Technology (DST) and J C Bose Grant ANRF/JBG/2026/000149/PS from Anusandhan National Research Foundation (ANRF), Government of India. SK would like to acknowledge the research support from the MATRICES Grant MTR/2023/000079, funded by SERB. Some of the computations are performed on HPC clusters procured with Swarna Jayanti Fellowship grants and the Core Research Grant CRG/2019/005373. SK acknowledges the Alexander von Humboldt Experienced Researcher Fellowship, which enabled SK and JH to initiate this collaborative work. JH acknowledges financial support from the DFG grant HO 2231/26-1 that enabled the visit of AB and SK to Heinrich Heine University D\"usseldorf, during which part of the work was conducted. AB acknowledges the prestigious Sarojini Damodaran International Student Travel Fellowship from TIFR.

\bibliography{activeHamiltonian}

@article{shee2020mapping,
	Author = {Shee, Amir and Dhar, Abhishek and Chaudhuri, Debasish},
	Doi = {10.1039/d0sm00367k},
	Eprint = {https://pubs.rsc.org/sm/article-pdf/16/20/4776/7567891/d0sm00367k.pdf},
	Issn = {1744-683X},
	Journal = {Soft Matter},
	Month = {05},
	Number = {20},
	Pages = {4776-4787},
	Title = {Active Brownian particles: mapping to equilibrium polymers and exact computation of moments},
	Url = {https://doi.org/10.1039/d0sm00367k},
	Volume = {16},
	Year = {2020}
    }

@article{RashmiNatComm2026,
	Author = {Priya, Rashmi and Horbach, J{\"u}rgen and Karmakar, Smarajit},
	Da = {2026/07/23},
	Doi = {10.1038/s41467-026-75709-y},
	Id = {Priya2026},
	Isbn = {2041-1723},
	Journal = {Nature Communications},
	Title = {Tunable yielding and emergent rheology in amorphous solids with active particle doping},
	Ty = {JOUR},
	Url = {https://doi.org/10.1038/s41467-026-75709-y},
	Year = {2026}}

@article{DeyNatComm2025,
  title={Enhanced long wavelength Mermin-Wagner-Hohenberg fluctuations in active crystals and glasses},
  author={Dey, Subhodeep and Bhattacharya, Antik and Karmakar, Smarajit},
  journal={Nature Communications},
  volume={16},
  number={1},
  pages={5498},
  year={2025},
  publisher={Nature Publishing Group UK London}
}

@article{sharmaNatPhys2025,
  title={Activity-induced annealing leads to a ductile-to-brittle transition in amorphous solids},
  author={Sharma, Rishabh and Karmakar, Smarajit},
  journal={Nature Physics},
  volume={21},
  number={2},
  pages={253--261},
  year={2025},
  publisher={Nature Publishing Group UK London}
}

@article{GoswamiNatPhys2025,
  title={Yielding behaviour of active particles in bulk and in confinement},
  author={Goswami, Yagyik and Shivashankar, GV and Sastry, Srikanth},
  journal={Nature Physics},
  volume={21},
  number={5},
  pages={817--824},
  year={2025},
  publisher={Nature Publishing Group UK London}
}

@article{Vicsek1995,
  title = {Novel Type of Phase Transition in a System of Self-Driven Particles},
  author = {Vicsek, Tam\'as and Czir\'ok, Andr\'as and Ben-Jacob, Eshel and Cohen, Inon and Shochet, Ofer},
  journal = {Phys. Rev. Lett.},
  volume = {75},
  issue = {6},
  pages = {1226--1229},
  numpages = {0},
  year = {1995},
  month = {Aug},
  publisher = {American Physical Society},
  doi = {10.1103/PhysRevLett.75.1226},
  url = {https://link.aps.org/doi/10.1103/PhysRevLett.75.1226}
}

@article{Fily2012,
  title = {Athermal Phase Separation of Self-Propelled Particles with No Alignment},
  author = {Fily, Yaouen and Marchetti, M. Cristina},
  journal = {Phys. Rev. Lett.},
  volume = {108},
  issue = {23},
  pages = {235702},
  numpages = {5},
  year = {2012},
  month = {Jun},
  publisher = {American Physical Society},
  doi = {10.1103/PhysRevLett.108.235702},
  url = {https://link.aps.org/doi/10.1103/PhysRevLett.108.235702}
}

@article{Bhattacharya2025,
  title = {Thermostatting of active Hamiltonian systems via symplectic algorithms},
  author = {Bhattacharya, Antik and Horbach, J\"urgen and Karmakar, Smarajit},
  journal = {Phys. Rev. E},
  volume = {111},
  issue = {1},
  pages = {015429},
  numpages = {17},
  year = {2025},
  month = {Jan},
  publisher = {American Physical Society},
  doi = {10.1103/PhysRevE.111.015429},
  url = {https://link.aps.org/doi/10.1103/PhysRevE.111.015429}
}

@article{Bhattacharya2026,
  title={Perspective: The Physics of Active Solids--From Hamiltonians to Active Matter Models},
  author={Bhattacharya, Antik and Horbach, J{\"u}rgen and Karmakar, Smarajit},
  journal={arXiv preprint arXiv:2606.11950},
  year={2026}
}

@article{Marchetti2013,
  title = {Hydrodynamics of soft active matter},
  author = {Marchetti, M. C. and Joanny, J. F. and Ramaswamy, S. and Liverpool, T. B. and Prost, J. and Rao, Madan and Simha, R. Aditi},
  journal = {Rev. Mod. Phys.},
  volume = {85},
  issue = {3},
  pages = {1143--1189},
  numpages = {0},
  year = {2013},
  month = {Jul},
  publisher = {American Physical Society},
  doi = {10.1103/RevModPhys.85.1143},
  url = {https://link.aps.org/doi/10.1103/RevModPhys.85.1143}
}

@article{Ramaswamy2010,
  author={Ramaswamy, S.},
  title={The Mechanics and Statistics of Active Matter},
  journal={Annu. Rev. Condens. Matter Phys.},
  volume={1},
  pages={323},
  year={2010}
}

@article{Netz2020,
  title = {Approach to equilibrium and nonequilibrium stationary distributions of interacting many-particle systems that are coupled to different heat baths},
  author = {Netz, Roland R.},
  journal = {Phys. Rev. E},
  volume = {101},
  issue = {2},
  pages = {022120},
  numpages = {23},
  year = {2020},
  month = {Feb},
  publisher = {American Physical Society},
  doi = {10.1103/PhysRevE.101.022120},
  url = {https://link.aps.org/doi/10.1103/PhysRevE.101.022120}
}

@article{Toner1998,
  title = {Flocks, herds, and schools: A quantitative theory of flocking},
  author = {Toner, John and Tu, Yuhai},
  journal = {Phys. Rev. E},
  volume = {58},
  issue = {4},
  pages = {4828--4858},
  numpages = {0},
  year = {1998},
  month = {Oct},
  publisher = {American Physical Society},
  doi = {10.1103/PhysRevE.58.4828},
  url = {https://link.aps.org/doi/10.1103/PhysRevE.58.4828}
}

@article{Fodor2016,
  title = {How Far from Equilibrium Is Active Matter?},
  author = {Fodor, \'Etienne and Nardini, Cesare and Cates, Michael E. and Tailleur, Julien and Visco, Paolo and van Wijland, Fr\'ed\'eric},
  journal = {Phys. Rev. Lett.},
  volume = {117},
  issue = {3},
  pages = {038103},
  numpages = {6},
  year = {2016},
  month = {Jul},
  publisher = {American Physical Society},
  doi = {10.1103/PhysRevLett.117.038103},
  url = {https://link.aps.org/doi/10.1103/PhysRevLett.117.038103}
}

@article{Bowick2022,
  title = {Symmetry, Thermodynamics, and Topology in Active Matter},
  author = {Bowick, Mark J. and Fakhri, Nikta and Marchetti, M. Cristina and Ramaswamy, Sriram},
  journal = {Phys. Rev. X},
  volume = {12},
  issue = {1},
  pages = {010501},
  numpages = {27},
  year = {2022},
  month = {Feb},
  publisher = {American Physical Society},
  doi = {10.1103/PhysRevX.12.010501},
  url = {https://link.aps.org/doi/10.1103/PhysRevX.12.010501}
}

@article{Cengio2019,
  title = {Linear Response Theory and Green-Kubo Relations for Active Matter},
  author = {Dal Cengio, Sara and Levis, Demian and Pagonabarraga, Ignacio},
  journal = {Phys. Rev. Lett.},
  volume = {123},
  issue = {23},
  pages = {238003},
  numpages = {6},
  year = {2019},
  month = {Dec},
  publisher = {American Physical Society},
  doi = {10.1103/PhysRevLett.123.238003},
  url = {https://link.aps.org/doi/10.1103/PhysRevLett.123.238003}
}

@article{solon2015pressure,
  title={Pressure is not a state function for generic active fluids},
  author={Solon, Alexandre P and Fily, Yaouen and Baskaran, Aparna and Cates, Mickael E and Kafri, Yariv and Kardar, Mehran and Tailleur, Julien},
  journal={Nature physics},
  volume={11},
  number={8},
  pages={673--678},
  year={2015},
  publisher={Nature Publishing Group UK London}
}

@article{solon2018generalized,
  title={Generalized thermodynamics of motility-induced phase separation: phase equilibria, Laplace pressure, and change of ensembles},
  author={Solon, Alexandre P and Stenhammar, Joakim and Cates, Michael E and Kafri, Yariv and Tailleur, Julien},
  journal={New Journal of Physics},
  volume={20},
  number={7},
  pages={075001},
  year={2018},
  publisher={IOP Publishing}
}

@article{Romanczuk2012,
  title={Active Brownian particles: From individual to collective stochastic dynamics},
  author={Romanczuk, Pawel and B{\"a}r, Markus and Ebeling, Werner and Lindner, Benjamin and Schimansky-Geier, Lutz},
  journal={The European Physical Journal Special Topics},
  volume={202},
  number={1},
  pages={1--162},
  year={2012},
  publisher={Springer}
}

@article{Bechinger2016,
  title = {Active particles in complex and crowded environments},
  author = {Bechinger, Clemens and Di Leonardo, Roberto and L\"owen, Hartmut and Reichhardt, Charles and Volpe, Giorgio and Volpe, Giovanni},
  journal = {Rev. Mod. Phys.},
  volume = {88},
  issue = {4},
  pages = {045006},
  numpages = {50},
  year = {2016},
  month = {Nov},
  publisher = {American Physical Society},
  doi = {10.1103/RevModPhys.88.045006},
  url = {https://link.aps.org/doi/10.1103/RevModPhys.88.045006}
}

@article{Szamel2014,
  title = {Self-propelled particle in an external potential: Existence of an effective temperature},
  author = {Szamel, Grzegorz},
  journal = {Phys. Rev. E},
  volume = {90},
  issue = {1},
  pages = {012111},
  numpages = {7},
  year = {2014},
  month = {Jul},
  publisher = {American Physical Society},
  doi = {10.1103/PhysRevE.90.012111},
  url = {https://link.aps.org/doi/10.1103/PhysRevE.90.012111}
}

@article{Fruchart2021,
  title={Non-reciprocal phase transitions},
  author={Fruchart, Michel and Hanai, Ryo and Littlewood, Peter B and Vitelli, Vincenzo},
  journal={Nature},
  volume={592},
  number={7854},
  pages={363--369},
  year={2021},
  publisher={Nature Publishing Group UK London}
}

@article{Casiulis2020,
  title = {Velocity and Speed Correlations in Hamiltonian Flocks},
  author = {Casiulis, Mathias and Tarzia, Marco and Cugliandolo, Leticia F. and Dauchot, Olivier},
  journal = {Phys. Rev. Lett.},
  volume = {124},
  issue = {19},
  pages = {198001},
  numpages = {6},
  year = {2020},
  month = {May},
  publisher = {American Physical Society},
  doi = {10.1103/PhysRevLett.124.198001},
  url = {https://link.aps.org/doi/10.1103/PhysRevLett.124.198001}
}

@article{Shi2026,
  title={Hamiltonian description of non-reciprocal interactions},
  author={Shi, Yu-Bo and Moessner, Roderich and Alert, Ricard and Bukov, Marin},
  journal={Nature Physics},
  pages={1--10},
  year={2026},
  publisher={Nature Publishing Group}
}

@incollection{risken1989fokker,
  title={Fokker-planck equation},
  author={Risken, Hannes},
  booktitle={The Fokker-Planck equation: methods of solution and applications},
  pages={63--95},
  year={1989},
  publisher={Springer}
}

@article{Shankar2018,
  title = {Hidden entropy production and work fluctuations in an ideal active gas},
  author = {Shankar, Suraj and Marchetti, M. Cristina},
  journal = {Phys. Rev. E},
  volume = {98},
  issue = {2},
  pages = {020604(R)},
  numpages = {6},
  year = {2018},
  month = {Aug},
  publisher = {American Physical Society},
  doi = {10.1103/PhysRevE.98.020604},
  url = {https://link.aps.org/doi/10.1103/PhysRevE.98.020604}
}

@article{kloeden1977numerical,
  title={The numerical solution of stochastic differential equations},
  author={Kloeden, Peter E and Pearson, RA},
  journal={The ANZIAM Journal},
  volume={20},
  number={1},
  pages={8--12},
  year={1977},
  publisher={Cambridge University Press}
}

@article{Howse2007,
  title = {Self-Motile Colloidal Particles: From Directed Propulsion to Random Walk},
  author = {Howse, Jonathan R. and Jones, Richard A. L. and Ryan, Anthony J. and Gough, Tim and Vafabakhsh, Reza and Golestanian, Ramin},
  journal = {Phys. Rev. Lett.},
  volume = {99},
  issue = {4},
  pages = {048102},
  numpages = {4},
  year = {2007},
  month = {Jul},
  publisher = {American Physical Society},
  doi = {10.1103/PhysRevLett.99.048102},
  url = {https://link.aps.org/doi/10.1103/PhysRevLett.99.048102}
}

@article{Pototsky2012,
  title={Active Brownian particles in two-dimensional traps},
  author={Pototsky, Andrey and Stark, Holger},
  journal={EPL (Europhysics Letters)},
  volume={98},
  number={5},
  pages={50004},
  year={2012}
}

@article{Malakar2020,
  title = {Steady state of an active Brownian particle in a two-dimensional harmonic trap},
  author = {Malakar, Kanaya and Das, Arghya and Kundu, Anupam and Kumar, K. Vijay and Dhar, Abhishek},
  journal = {Phys. Rev. E},
  volume = {101},
  issue = {2},
  pages = {022610},
  numpages = {10},
  year = {2020},
  month = {Feb},
  publisher = {American Physical Society},
  doi = {10.1103/PhysRevE.101.022610},
  url = {https://link.aps.org/doi/10.1103/PhysRevE.101.022610}
}

@article{Nakul2023,
  title = {Stationary states of an active Brownian particle in a harmonic trap},
  author = {Nakul, Urvashi and Gopalakrishnan, Manoj},
  journal = {Phys. Rev. E},
  volume = {108},
  issue = {2},
  pages = {024121},
  numpages = {15},
  year = {2023},
  month = {Aug},
  publisher = {American Physical Society},
  doi = {10.1103/PhysRevE.108.024121},
  url = {https://link.aps.org/doi/10.1103/PhysRevE.108.024121}
}

@book{hansen2013theory,
  title={Theory of simple liquids: with applications to soft matter},
  author={Hansen, Jean-Pierre and McDonald, Ian Ranald},
  year={2013},
  publisher={Academic press}
}

@article{Seifert2012,
  title={Stochastic thermodynamics, fluctuation theorems and molecular machines},
  author={Seifert, Udo},
  journal={Reports on progress in physics},
  volume={75},
  number={12},
  pages={126001},
  year={2012},
  publisher={IOP Publishing}
}

@article{Filliger2007,
  title = {Brownian Gyrator: A Minimal Heat Engine on the Nanoscale},
  author = {Filliger, Roger and Reimann, Peter},
  journal = {Phys. Rev. Lett.},
  volume = {99},
  issue = {23},
  pages = {230602},
  numpages = {4},
  year = {2007},
  month = {Dec},
  publisher = {American Physical Society},
  doi = {10.1103/PhysRevLett.99.230602},
  url = {https://link.aps.org/doi/10.1103/PhysRevLett.99.230602}
}

@article{Speck2016,
  title={Stochastic thermodynamics for active matter},
  author={Speck, Thomas},
  journal={Europhysics Letters},
  volume={114},
  number={3},
  pages={30006},
  year={2016},
  publisher={EDP Sciences, IOP Publishing and Societ{\`a} Italiana di Fisica}
}

@article{Pietzonka2018,
  title={Entropy production of active particles and for particles in active baths},
  author={Pietzonka, Patrick and Seifert, Udo},
  journal={Journal of Physics A: Mathematical and Theoretical},
  volume={51},
  number={1},
  pages={01LT01},
  year={2018},
  publisher={IOP Publishing}
}

@article{Mandal2017,
  title = {Entropy Production and Fluctuation Theorems for Active Matter},
  author = {Mandal, Dibyendu and Klymko, Katherine and DeWeese, Michael R.},
  journal = {Phys. Rev. Lett.},
  volume = {119},
  issue = {25},
  pages = {258001},
  numpages = {6},
  year = {2017},
  month = {Dec},
  publisher = {American Physical Society},
  doi = {10.1103/PhysRevLett.119.258001},
  url = {https://link.aps.org/doi/10.1103/PhysRevLett.119.258001}
}

@article{Casiulis2026,
  title={Hamiltonian flocks: time-reversal symmetry and its consequences},
  author={Casiulis, Mathias and Cugliandolo, Leticia F},
  journal={Journal of Statistical Mechanics: Theory and Experiment},
  volume={2026},
  number={7},
  pages={073207},
  year={2026},
  publisher={IOP Publishing}
}

@article{Liebchen2018,
  title={Synthetic chemotaxis and collective behavior in active matter},
  author={Liebchen, Benno and Lowen, Hartmut},
  journal={Accounts of chemical research},
  volume={51},
  number={12},
  pages={2982--2990},
  year={2018},
  publisher={ACS Publications}
}

@article{Martin2021,
  title = {Statistical mechanics of active Ornstein-Uhlenbeck particles},
  author = {Martin, David and O'Byrne, J\'er\'emy and Cates, Michael E. and Fodor, \'Etienne and Nardini, Cesare and Tailleur, Julien and van Wijland, Fr\'ed\'eric},
  journal = {Phys. Rev. E},
  volume = {103},
  issue = {3},
  pages = {032607},
  numpages = {25},
  year = {2021},
  month = {Mar},
  publisher = {American Physical Society},
  doi = {10.1103/PhysRevE.103.032607},
  url = {https://link.aps.org/doi/10.1103/PhysRevE.103.032607}
}

@article{Solon2015,
  title = {Pressure and Phase Equilibria in Interacting Active Brownian Spheres},
  author = {Solon, Alexandre P. and Stenhammar, Joakim and Wittkowski, Raphael and Kardar, Mehran and Kafri, Yariv and Cates, Michael E. and Tailleur, Julien},
  journal = {Phys. Rev. Lett.},
  volume = {114},
  issue = {19},
  pages = {198301},
  numpages = {6},
  year = {2015},
  month = {May},
  publisher = {American Physical Society},
  doi = {10.1103/PhysRevLett.114.198301},
  url = {https://link.aps.org/doi/10.1103/PhysRevLett.114.198301}
}

@article{Digregorio2018,
  title = {Full Phase Diagram of Active Brownian Disks: From Melting to Motility-Induced Phase Separation},
  author = {Digregorio, Pasquale and Levis, Demian and Suma, Antonio and Cugliandolo, Leticia F. and Gonnella, Giuseppe and Pagonabarraga, Ignacio},
  journal = {Phys. Rev. Lett.},
  volume = {121},
  issue = {9},
  pages = {098003},
  numpages = {5},
  year = {2018},
  month = {Aug},
  publisher = {American Physical Society},
  doi = {10.1103/PhysRevLett.121.098003},
  url = {https://link.aps.org/doi/10.1103/PhysRevLett.121.098003}
}

@article{Buttinoni2013,
  title = {Dynamical Clustering and Phase Separation in Suspensions of Self-Propelled Colloidal Particles},
  author = {Buttinoni, Ivo and Bialk\'e, Julian and K\"ummel, Felix and L\"owen, Hartmut and Bechinger, Clemens and Speck, Thomas},
  journal = {Phys. Rev. Lett.},
  volume = {110},
  issue = {23},
  pages = {238301},
  numpages = {5},
  year = {2013},
  month = {Jun},
  publisher = {American Physical Society},
  doi = {10.1103/PhysRevLett.110.238301},
  url = {https://link.aps.org/doi/10.1103/PhysRevLett.110.238301}
}

@article{Cavagna2014,
  title={Bird flocks as condensed matter},
  author={Cavagna, Andrea and Giardina, Irene},
  journal={Annu. Rev. Condens. Matter Phys.},
  volume={5},
  number={1},
  pages={183--207},
  year={2014},
  publisher={Annual Reviews}
}

@article{Cavagna2017,
  title={Dynamic scaling in natural swarms},
  author={Cavagna, Andrea and Conti, Daniele and Creato, Chiara and Del Castello, Lorenzo and Giardina, Irene and Grigera, Tomas S and Melillo, Stefania and Parisi, Leonardo and Viale, Massimiliano},
  journal={Nature Physics},
  volume={13},
  number={9},
  pages={914--918},
  year={2017},
  publisher={Nature Publishing Group UK London}
}

@article{Chen2026,
  title={Reentrance in a Hamiltonian flocking model},
  author={Chen, Letian and Davis, Luke K},
  journal={arXiv preprint arXiv:2602.11104},
  year={2026}
}

@article{Caraglio2022,
  title = {Analytic Solution of an Active Brownian Particle in a Harmonic Well},
  author = {Caraglio, Michele and Franosch, Thomas},
  journal = {Phys. Rev. Lett.},
  volume = {129},
  issue = {15},
  pages = {158001},
  numpages = {8},
  year = {2022},
  month = {Oct},
  publisher = {American Physical Society},
  doi = {10.1103/PhysRevLett.129.158001},
  url = {https://link.aps.org/doi/10.1103/PhysRevLett.129.158001}
}

@article{Fieguth2022,
  title = {Hamiltonian active particles in an environment},
  author = {Fieguth, Diego M. and Schlachter, Timo and Brady, Daniel S. and Anglin, James R.},
  journal = {Phys. Rev. E},
  volume = {106},
  issue = {4},
  pages = {044201},
  numpages = {13},
  year = {2022},
  month = {Oct},
  publisher = {American Physical Society},
  doi = {10.1103/PhysRevE.106.044201},
  url = {https://link.aps.org/doi/10.1103/PhysRevE.106.044201}
}

@article{Fieguth2024,
  title={Dynamical active particles in the overdamped limit},
  author={Fieguth, Diego M},
  journal={Journal of Physics Communications},
  volume={8},
  number={7},
  pages={075001},
  year={2024},
  publisher={IOP Publishing}
}

@article{Chaudhuri2021,
  title={Active Brownian particle in harmonic trap: exact computation of moments, and re-entrant transition},
  author={Chaudhuri, Debasish and Dhar, Abhishek},
  journal={Journal of Statistical Mechanics: Theory and Experiment},
  volume={2021},
  number={1},
  pages={013207},
  year={2021},
  publisher={IOP Publishing and SISSA}
}

@article{Sekimoto1998,
  title={Langevin equation and thermodynamics},
  author={Sekimoto, Ken},
  journal={Progress of Theoretical Physics Supplement},
  volume={130},
  pages={17--27},
  year={1998},
  publisher={Oxford Academic}
}

\onecolumngrid
\begin{center}
    \vspace{0.5cm}
    {\large\bf End Matter}
    \vspace{0.3cm}
\end{center}
\twocolumngrid

The basis for the calculations presented in Appendices A and B is the Hamilton function for the two-dimensional HABP model, given by
\begin{equation}
\mathcal{H} = \frac{\vec{p}^{\, 2}}{2m} + \frac{L^2}{2I} -  K \vec{S} \cdot \vec{r}
+ \frac{\lambda}{2} \vec{r}^{\, 2} \, ,
\label{eq_em1} 
\end{equation} 
where we have included a harmonic trap potential with trap strength $\lambda$. The corresponding overdamped Langevin equations for the HABP model with a coupling to two temperature baths with temperatures $T_{\textrm{tr}}$ and $T_\theta$ are
\begin{eqnarray}
\gamma_{\textrm{tr}} \dot{\vec{r}} & = & 
- \frac{\partial \mathcal{H}}{\partial \vec{r}} + \sqrt{2k_{\textrm{B}}T_{\textrm{tr}}\gamma_{\textrm{tr}}} \;  
\vec{\xi}(t) , \label{eq_em2} \\
\gamma_\theta \dot{\theta} & = & 
- \frac{\partial \mathcal{H}}{\partial \theta}+ 
\sqrt{2k_{\textrm{B}}T_{\theta}\gamma_{\theta}} \; \eta(t), 
\label{eq_em3}
\end{eqnarray}
where $- \frac{\partial \mathcal{H}}{\partial \vec{r}} = K \vec{S} - \lambda \vec{r}$ and $-\frac{\partial \mathcal{H}}{\partial \theta}= K \vec{S}^\perp \cdot \vec{r}$.

%
%

{\bf Appendix A: Plateau height of the translational MSD for the HABP in a harmonic trap at equilibrium.} In this appendix, we derive the analytical solution of the long-time behavior of the translational MSD for a HABP in a harmonic trap at equilibrium ($T=T_\theta = T_{\textrm{tr}}$). In this case, the particle is localized in the harmonic trap and thus at long times the MSD reaches the constant $\langle \Delta r^2_{\textrm{plat}} \rangle$. To compute this constant, we consider the configurational part of the Hamilton function \eqref{eq_em1} which is $\mathcal{H}_{\textrm{conf}} = -  K \vec{S} \cdot \vec{r} + \frac{\lambda}{2} \vec{r}^{\, 2}$; $\mathcal{H}_{\textrm{conf}}$ can be exactly rewritten in the following form:
\begin{equation}
\mathcal{H}_{\textrm{conf}} = \frac{\lambda}{2} \left| \vec{r} - \frac{K}{\lambda} \vec{S} \right|^2 - \frac{K^2}{2\lambda}
\label{eq_em4}
\end{equation}
At equilibrium, the stationary distribution of the HABP in the harmonic trap is a Boltzmann distribution, $\rho_{\textrm{eq}} \propto \exp(-\mathcal{H}_{\textrm{conf}}/(k_{\textrm{B}} T))$. Using Eq.~\eqref{eq_em4}, we can write
\begin{equation}
\rho_{\textrm{eq}} \propto \exp\left(  \frac{\lambda}{2} \left| \vec{r} - \frac{K}{\lambda} \vec{S} \right|^2  \right) \, .
\label{eq_em5}
\end{equation}
For a fixed $\theta$, i.e.~a fixed spin $\vec{S}$, the distribution of the position $\vec{r}$ is a Gauss function centered at $\vec{r}_0(\theta) = \frac{K}{\lambda} \vec{S}(\theta)$ with variance $\frac{k_{\textrm{B}}T}{\lambda}$ per Cartesian component. The second moment of the Gauss distribution corresponds to plateau height of the translational MSD, that we are looking for. It is given by  
\begin{equation}
\langle \Delta r^2_{\textrm{plat}}\rangle = \vec{r}_0^{\, 2} + \frac{2 k_{\textrm{B}} T}{\lambda} 
= \frac{K^2}{\lambda^2}  + \frac{2 k_{\textrm{B}} T}{\lambda} \, .
\label{eq_em6}
\end{equation}
Here, we have used $\vec{S}^2 =1$; as a consequence $\langle \Delta r^2_{\textrm{plat}}\rangle$ is independent of $\theta$.

{\bf Appendix B: Entropy production of the HABP model.} We now compute the entropy production of the HABP under non-equilibrium conditions. Based on the Langevin equations \eqref{eq_em2} and \eqref{eq_em3}, the heat currents into the translational and rotational baths are obtained from the work performed by the corresponding dissipative channels \cite{Sekimoto1998, Seifert2012}. For the instantaneous heat rates, $\dot{Q}_{\textrm{tr}}$ in the translational sector and $\dot{Q}_\theta$ in the rotational sector, we find
\begin{eqnarray}
\dot{Q}_{\textrm{tr}} & = &
- \frac{\partial \mathcal{H}}{\partial \vec{r}} \cdot \dot{\vec{r}}
= (K\vec{S} - \lambda \vec{r}) \cdot \dot{\vec{r}}, \label{eq_em7} \\
\dot{Q}_\theta & = &
- \frac{\partial \mathcal{H}}{\partial \theta}\,\dot{\theta}
= K (\vec{S}^{\perp} \cdot \vec{r})\,\dot{\theta} \, .
\label{eq_em8}
\end{eqnarray}
Note that these formulas implicitly assume the use of Stratonovich calculus.

Differentiating the Hamiltonian along a stochastic trajectory gives
\begin{equation}
\dot{\mathcal{H}} =
\frac{\partial \mathcal{H}}{\partial \vec{r}} \cdot \dot{\vec{r}}
+ \frac{\partial \mathcal{H}}{\partial \theta} \, \dot{\theta}.
\label{eq_em9}
\end{equation}
Using Eqs.~\eqref{eq_em7} and \eqref{eq_em8},
\begin{equation}
\dot{\mathcal{H}} = - (\dot{Q}_{\textrm{tr}} + \dot{Q}_\theta).
\label{eq_em10}
\end{equation}
In the steady state,
\begin{equation}
\langle \dot{\mathcal{H}} \rangle = 0 \quad
\Rightarrow \quad
\langle \dot{Q}_{\textrm{tr}} \rangle + \langle \dot{Q}_\theta \rangle = 0.
\label{eq_em11}
\end{equation}
Thus, in the steady state, the spin-position coupling mediates a heat current between the two baths without net energy accumulation.

From the heat rates, the steady-state entropy production $\dot{\sigma}$ can be obtained as
\begin{equation}
\dot{\sigma} = \frac{1}{k_{\textrm{B}}}
\left\langle \frac{\dot{Q}_{\textrm{tr}}}{T_{\textrm{tr}}} 
+ \frac{\dot{Q}_\theta}{T_\theta} \right\rangle.
\label{eq_em12}
\end{equation}
Using Eq.~\eqref{eq_em11}, this becomes
\begin{equation}
\dot{\sigma} = \frac{1}{k_{\textrm{B}}}
\left(\frac{1}{T_{\textrm{tr}}} - \frac{1}{T_\theta}\right) 
\langle \dot{Q}_{\textrm{tr}} \rangle = \frac{1}{k_{\textrm{B}}}
\left(\frac{1}{T_\theta} - \frac{1}{T_{\textrm{tr}}} \right)
\langle \dot{Q}_\theta \rangle.
\label{eq_em13}
\end{equation}
This expression is exact and holds for both $T_\theta > T_{\textrm{tr}}$ and $T_\theta < T_{\textrm{tr}}$. The sign of the heat current reverses when the temperature ordering is reversed, so that $\dot{\sigma}\ge 0$ in all cases.

To obtain $\langle \dot{Q}_{\textrm{tr}} \rangle$, we take the average of Eq.~\eqref{eq_em7},
\begin{equation}
\langle \dot{Q}_{\textrm{tr}} \rangle =
K \langle \vec{S} \cdot \dot{\vec{r}} \rangle -
\lambda \langle \vec{r} \cdot \dot{\vec{r}} \rangle =
K \langle \vec{S} \cdot \dot{\vec{r}} \rangle.
\label{eq_em14}
\end{equation}
since $\langle \vec{r} \cdot \dot{\vec{r}} \rangle = 0$ in the steady state. Now take the scalar product of Eq.~\eqref{eq_em2} with $\vec{S}$:
\begin{equation}
\gamma_{\textrm{tr}} \vec{S} \cdot \dot{\vec{r}} =
K \vec{S} \cdot \vec{S} - \lambda \vec{S} \cdot \vec{r}
+ \sqrt{2 k_{\textrm{B}} T_{\textrm{tr}} \gamma_{\textrm{tr}}}\,
\vec{S} \cdot \vec{\xi}(t) \, .
\label{eq_em15}
\end{equation}
Averaging over the steady state, the noise term vanishes, and since $\vec{S} \cdot \vec{S} = 1$ we obtain
\begin{equation}
\gamma_{\textrm{tr}} \langle \vec{S} \cdot \dot{\vec{r}} \rangle
= K - \lambda \langle \vec{S} \cdot \vec{r} \rangle \, .
\label{eq_em16}
\end{equation}
Thus, we obtain
\begin{equation}
\langle \dot{Q}_{\textrm{tr}} \rangle =
\frac{K^2}{\gamma_{\textrm{tr}}} -
\frac{\lambda K}{\gamma_{\textrm{tr}}}\,
\langle \vec{S} \cdot \vec{r} \rangle \, .
\label{eq_em17}
\end{equation}
Substituting Eq.~\eqref{eq_em17} into Eq.~\eqref{eq_em13} yields the entropy production rate:
\begin{equation}
\dot{\sigma} = \frac{1}{k_{\textrm{B}}} \left(\frac{1}{T_{\textrm{tr}}} - \frac{1}{T_\theta} \right)
\left[ \frac{K^2}{\gamma_{\textrm{tr}}} -
\frac{\lambda K}{\gamma_{\textrm{tr}}}\,
\langle \vec{S} \cdot \vec{r} \rangle \right] \, .
\label{eq_em18}
\end{equation}

For the free particle, $\lambda=0$, Eq.~\eqref{eq_em17} reduces to
\begin{equation}
\langle \dot{Q}_{\textrm{tr}} \rangle = \frac{K^2}{\gamma_{\textrm{tr}}} \, ,
\label{eq_em19}
\end{equation}
and thus the entropy production rate is 
\begin{equation}
\dot{\sigma} = \frac{K^2}{\gamma_{\textrm{tr}} k_{\textrm{B}}}
\left(\frac{1}{T_\textrm{tr}} - \frac{1}{T_\theta} \right) \, .
\label{eq_em20}
\end{equation}
Defining the propulsion speed $v_0 = K/\gamma_{\textrm{tr}}$, this can be written as
\begin{equation}
\dot{\sigma} = \frac{\gamma_{\textrm{tr}} v_0^2}{k_{\textrm{B}}} \left(\frac{1}{T_\textrm{tr}} - \frac{1}{T_\theta} \right) \, ,
\label{eq_em21}
\end{equation}
which reduces for $T_\theta \to \infty$ to the known ABP result for TRS-even propulsion \cite{Shankar2018}, $\dot{\sigma} = \frac{\gamma_{\textrm{tr}} v_0^2}{k_{\textrm{B}}T_{\textrm{tr}}}$.

For the HABP in a harmonic trap, we have to evaluate $\langle \vec{S} \cdot \vec{r} \rangle$. An analytical solution can be obtained for $T_\theta \gg T_{\textrm{tr}}$. Then, the rotational dynamics is noise-dominated and so to leading order one can neglect the feedback torque in the angular equation and write
\begin{equation}
\dot{\theta} = \sqrt{2 D_\theta} \eta(t) \, ,
\label{eq_em22}
\end{equation}
with $D_\theta = \frac{k_{\textrm{B}} T_\theta}{\gamma_\theta}$. In this case, $\theta(t)$ is a Wiener process on the circle, and the orientational correlation is
\begin{equation}
\langle \vec{S}(t) \cdot \vec{S}(t^\prime) \rangle \approx {\rm e}^{-D_\theta |t-t^\prime|} \, .
\label{eq_em23}
\end{equation}
The translational equation \eqref{eq_em2} can be rewritten as 
\begin{equation}
\dot{\vec{r}} + \frac{\lambda}{\gamma_{\textrm{tr}}} \vec{r} = 
\frac{K}{\gamma_{\textrm{tr}}} \vec{S} + \sqrt{\frac{2k_{\textrm{B}} T_{\textrm{tr}}}{\gamma_{\textrm{tr}}}} \vec{\xi}(t) \, .
\label{eq_em24}
\end{equation}
This equation has the stationary solution
\begin{eqnarray}
\vec{r}(t) & = & \frac{K}{\gamma_{\textrm{tr}}} \int_{-\infty}^t \, 
{\rm e}^{-\lambda (t-t^\prime)/\gamma_{\textrm{tr}}}\vec{S}(t^\prime)\mathrm{d}t^\prime \nonumber \\
& + & \sqrt{\frac{2 k_{\textrm{B}} T_{\textrm{tr}}}{\gamma_{\textrm{tr}}}} 
\int_{-\infty}^t \, {\rm e}^{-\lambda (t- t^\prime)/\gamma_{\textrm{tr}}} \vec{\xi}(t) \mathrm{d} t^\prime
\label{eq_em25}
\end{eqnarray}
Substitution with $s=t-t^\prime$, then taking the dot product with $\vec{S}$ and performing a stationary average yield
\begin{equation}
\langle \vec{S} \cdot \vec{r} \rangle = \frac{K}{\gamma_{\textrm{tr}}} \int_0^\infty {\rm e}^{-\lambda s/\gamma_{\textrm{tr}}}
\langle \vec{S}(t) \cdot \vec{S}(t-s) \rangle \, \mathrm{d} s \, .
\label{eq_em26}
\end{equation}
To obtain Eq.~\eqref{eq_em26}, we have used $\langle \vec{S}(t) \cdot \vec{\xi}(t-s) \rangle = 0$. Since $\langle \vec{S}(t) \cdot \vec{S}(t-s) \rangle = {\rm e}^{-D_\theta s}$ [cf.~Eq.~\eqref{eq_em23}], we obtain 
\begin{equation}
\langle \vec{S} \cdot \vec{r} \rangle = 
\frac{K}{\gamma_{\textrm{tr}}} \int_0^\infty 
\exp\left\{ - \left( \frac{\lambda}{\gamma_{\textrm{tr}}} +D_\theta \right) s \right\} \mathrm{d} s \, ,
\label{eq_em27}
\end{equation}
and hence
\begin{equation}
\langle \vec{S} \cdot \vec{r} \rangle = \frac{K}{\lambda + \gamma_{\textrm{tr}} D_\theta} \, .
\label{eq_em28}
\end{equation}
Inserting this result in Eq.~\eqref{eq_em18}, we obtain
\begin{equation}
\dot{\sigma} = \frac{1}{k_{\textrm{B}}} \left(\frac{1}{T_{\textrm{tr}}} - \frac{1}{T_\theta} \right)
\frac{K^2 D_\theta}{\lambda + \gamma_{\textrm{tr}} D_\theta} \, .
\label{eq_em29}
\end{equation}
In the derivation of Eq.~\eqref{eq_em29}, we have neglected the torque term $K \vec{S}^\perp \cdot \vec{r}$. This is valid when the orientational diffusion is fast enough that the back-coupling of $\vec{r}$ on $\theta$ is weak on the relevant time scale. A rough condition is $K |\vec{r}| \ll \gamma_\theta D_\theta \sim k_{\textrm{B}} T_\theta$. Using the typical displacement of the particle in the trap, $|\vec{r}| \sim K/(\lambda + \gamma_{\textrm{tr}} D_\theta)$, this becomes
\begin{equation}
\frac{K^2}{\lambda + \gamma_{\textrm{tr}} D_\theta} \ll k_{\textrm{B}} T_\theta \, .
\label{eq_em30}
\end{equation}
This condition defines the regime where Eq.~\eqref{eq_em29} is valid.

\end{document}


\title{Active Brownian Dynamics from a Hamiltonian Model - Supplemental Material}

\author{Antik Bhattacharya}
\affiliation{Tata Institute of Fundamental Research Hyderabad, 36/P, Gopanpally Village, Serilingampally Mandal, Ranga Reddy District, Hyderabad, Telangana 500046, India}
\author{Smarajit Karmakar}
\affiliation{Tata Institute of Fundamental Research Hyderabad, 36/P, Gopanpally Village, Serilingampally Mandal, Ranga Reddy District, Hyderabad, Telangana 500046, India}
\author{J\"urgen Horbach}
\affiliation{Institut f\"ur Theoretische Physik II: Weiche Materie, Heinrich-Heine-Universit\"at D\"usseldorf, Universit\"atsstra\ss e 1, 40225 D\"usseldorf, Germany}

\begin{abstract}
In the following, we present a theoretical derivation for the long-time plateau height of the rotational mean-square displacement (MSD) for a free Hamiltonian Active Brownian Particle (HABP) in equilibrium.
\end{abstract}

\maketitle

The Hamilton function for a free HABP in two dimensions is given by
\begin{equation}
\mathcal{H} = 
\frac{\vec{p}^{\, 2}}{2m} + \frac{L^2}{2I} 
-  K \vec{S} \cdot \vec{r} \, ,
\label{eq_sm1} 
\end{equation} 
with the coupling constant $K>0$. For the definitions of all variables in Eq.~\eqref{eq_sm1}, we refer the reader to the main manuscript. The corresponding overdamped Langevin equations for the HABP model, coupled to two heat baths at temperatures $T_{\textrm{tr}}$ and $T_\theta$, can be written as
\begin{eqnarray}
\gamma_{\textrm{tr}} \dot{\vec{r}} & = & 
K \vec{S} 
+ \sqrt{2k_{\textrm{B}}T_{\textrm{tr}}\gamma_{\textrm{tr}}} \;  
\vec{\xi}(t) \, , \label{eq_sm2} \\
\gamma_\theta \dot{\theta} & = & 
K \vec{S}^\perp \cdot \vec{r} +
\sqrt{2k_{\textrm{B}}T_{\theta}\gamma_{\theta}} \; \eta(t) \, . 
\label{eq_sm3}
\end{eqnarray}
Here, we consider a free HABP in equilibrium where the temperatures of the two baths are equal, $T= T_{\textrm{tr}} = T_\theta$. Equations \eqref{eq_sm2} and \eqref{eq_sm3} form the basis for calculating the plateau height $\langle \Delta \theta^2_{\textrm{plat}} \rangle$ that appears as the long-time limit of the rotational MSD, $\langle \Delta \theta^2(t) \rangle = \langle (\theta(t) - \theta(0))^2 \rangle$.

To this end, we first rewrite the angular equation \eqref{eq_sm3} in polar coordinates, $\vec{r} = r (\cos(\phi), \sin(\phi))^{\textrm{T}}$ where $\phi$ is the polar angle ($0\le \phi \le 2 \pi$). Using $\vec{S}^\perp \cdot \vec{r} = r \sin(\psi)$ (with $\psi = \phi - \theta$), we obtain
\begin{equation}
\gamma_\theta \dot{\theta} =
K r \sin(\psi)
+ \sqrt{2k_{\textrm{B}}T \gamma_{\theta}} \; \eta(t) \, .
\label{eq_sm4}
\end{equation}
The translational equation \eqref{eq_sm2} can also be written in polar coordinates. In the polar basis, we have $\dot{\vec{r}} = \dot{r} \hat{\vec{e}}_r + r \dot{\phi} \hat{\vec{e}}_\phi$ (with the unit vectors  $\hat{\vec{e}}_r = (\cos(\phi), \sin(\phi))^{\textrm{T}}$ and  $\hat{\vec{e}}_\phi = (- \sin(\phi), \cos(\phi))^{\textrm{T}}$) and $\vec{S} = \cos(\psi) \hat{\vec{e}}_r + \sin(\psi) \hat{\vec{e}}_\phi$. Thus, the translational equations of motion are
\begin{eqnarray}
\gamma_{\textrm{tr}} \dot{r} & = & 
K \cos(\psi)
+ \sqrt{2k_{\textrm{B}}T\gamma_{\textrm{tr}}} \;  
\xi_r(t) , \label{eq_sm5} \\
\gamma_{\textrm{tr}} r \dot{\phi} & = & 
K \sin(\psi)
+ \sqrt{2k_{\textrm{B}}T\gamma_{\textrm{tr}}} \;  
\xi_\phi(t) , \label{eq_sm6}
\end{eqnarray}
where $\xi_r(t)$ and $\xi_\phi(t)$ are the noise components in the polar basis.

For sufficiently long times, the relative angle $\psi$ remains bounded and exhibits small fluctuations $\delta \psi$ around the locked value $\psi_0$. Thus, we can write $\sin(\psi) = \sin(\psi_0 + \delta \psi) \approx \sin(\psi_0) + \cos(\psi_0) \delta \psi + \mathcal{O}(\delta \psi^2)$ and $\cos(\psi) \approx \cos(\psi_0) + \mathcal{O}(\delta \psi^2)$. Thus, neglecting terms $\mathcal{O}(\delta \psi^2)$, Eqs.~\eqref{eq_sm4}-\eqref{eq_sm6} become
\begin{eqnarray}
\gamma_\theta \dot{\theta} & \approx & 
K r \cos(\psi_0)\, \delta \psi 
+ \sqrt{2k_{\textrm{B}}T \gamma_{\theta}} \; \eta(t) \, ,
\label{eq_sm7} \\
\gamma_{\textrm{tr}} \dot{r} & = & 
K \cos(\psi_0)
+ \sqrt{2k_{\textrm{B}}T\gamma_{\textrm{tr}}} \;  
\xi_r(t) \, , \label{eq_sm8} \\
\gamma_{\textrm{tr}} r \dot{\phi} & = & 
K \cos(\psi_0) \delta \psi
+ \sqrt{2k_{\textrm{B}}T\gamma_{\textrm{tr}}} \;  
\xi_\phi(t) \, . \label{eq_sm9}
\end{eqnarray}
A stable locked state requires that $\sin(\psi_0) = 0$ (cf.~Eq.~\eqref{eq_sm6})  and thus $\psi_0 = 0, \pi$, and $\cos(\psi_0)$ can take the two possible values $+1$ and $-1$.  However, with the constraint that the speed $\dot{r}$ in Eq.~\eqref{eq_sm8} has to be non-negative, one has to choose $\psi_0 = 0$ such that $\cos(\psi_0)=1$.

Integrating Eq.~\eqref{eq_sm8} (with $\cos(\psi_0)=1$) yields
\begin{equation}
r(t) = r(0) + \frac{K}{\gamma_{\textrm{tr}}} t 
+ \sqrt{2 k_{\textrm{B}}T/\gamma_{\textrm{tr}}} 
\int_0^t \xi_r(s) \mathrm{d}s \, .
\label{eq_sm10a}
\end{equation}
We assume that $r(0) = 0$. Then, we obtain a deterministic term, $r_{\textrm{det}} = \frac{K}{\gamma_{\textrm{tr}}} t$, that indicates the ballistic motion of the particle, and a noise term with standard deviation $\sigma_r = \sqrt{2 k_{\textrm{B}}T t/\gamma_{\textrm{tr}}}$. The ratio $\sigma_r/r_{\textrm{det}}$ is proportional to $1/\sqrt{t}$ and vanishes in the limit $t\to \infty$. The crossover time scale, $\tau_{\textrm{tr}}$, at which the ballistic motion starts to dominate, can be obtained from the condition $\sigma_r(\tau_{\textrm{tr}}) \sim r_{\textrm{det}}(\tau_{\textrm{tr}})$ and is hence given by $\tau_{\textrm{tr}} = 4 k_{\textrm{B}}T \gamma_{\textrm{tr}}/K^2$. Thus, for $t\gg \tau_{\textrm{tr}}$, we may neglect the radial noise $\propto \xi_r$ and we can write
\begin{equation}
r(t) \approx \frac{K}{\gamma_{\textrm{tr}}} \, t \, .
\label{eq_sm10}
\end{equation}
That the use of this approximation is justified in the following can be inferred from Fig.~1 of the main manuscript, which demonstrates that $\tau_{\textrm{tr}}$ is significantly smaller than the time scale $\tau_{\theta}$ above which one observes the angular locking in the rotational MSD.

With the approximation \eqref{eq_sm10}, Eqs.~\eqref{eq_sm7} and \eqref{eq_sm9} are given by
\begin{eqnarray}
\dot{\theta} & = & \frac{K^2 t}{\gamma_\theta \gamma_{\textrm{tr}}} \delta \psi
+ \sqrt{\frac{2 k_{\textrm{B}} T}{\gamma_\theta}} \eta(t) \, , \label{eq_sm11} \\
\dot{\phi} & = & \frac{1}{t} \delta \psi 
+ \sqrt{\frac{2k_{\textrm{B}}T}{\gamma_{\textrm{tr}}}} \xi_\phi(t) \, .
\label{eq_sm12}
\end{eqnarray}
Subtracting Eq.~\eqref{eq_sm11} from \eqref{eq_sm12} yields
\begin{eqnarray}
\dot{\psi} & = & \dot{\delta \psi} = \dot{\phi} - \dot{\theta} \nonumber \\
& = & \left( \frac{1}{t} - \frac{K^2 t}{\gamma_\theta \gamma_{\textrm{tr}}} \right) 
\delta \psi \nonumber \\
& & + \sqrt{\frac{2k_{\textrm{B}}T}{\gamma_{\textrm{tr}}}} \xi_\phi(t)
- \sqrt{\frac{2 k_{\textrm{B}} T}{\gamma_\theta}} \eta(t) \, . 
\label{eq_sm13} 
\end{eqnarray}
For large times $t$, we can neglect the term $\delta \psi/t$ and the corresponding noise term and thus with the definitions $\kappa = \frac{K^2}{\gamma_\theta \gamma_{\textrm{tr}}}$ and $D_\theta = \frac{k_{\textrm{B}} T}{\gamma_\theta}$ we obtain the following equation:
\begin{equation}
\dot{\delta \psi} = 
- \kappa t  \delta \psi
- \sqrt{2 D_\theta} \eta(t) \, . 
\label{eq_sm14} 
\end{equation}
This is an Ornstein-Uhlenbeck equation for $\delta \psi$ with time-dependent  stiffness $\gamma_\theta \kappa t$ \cite{Risken1989}. The solution to Eq.~\eqref{eq_sm14} can be written as \cite{Risken1989} 
\begin{equation}
\delta \psi(t) = \delta \psi(0) {\rm e}^{-\kappa t^2}
+ \sqrt{2 D_\theta} \int_0^t {\rm e}^{-\kappa (t^2 - s^2)/2} \mathrm{d} s \, .
\label{eq_sm15}
\end{equation}
For long time, the first term ${\rm e}^{-\kappa t^2}$ can be neglected; then, squaring Eq.~\eqref{eq_sm15} and taking the thermal average $\langle \dots \rangle$ yield the variance 
\begin{equation}
\langle \delta \psi^2(t) \rangle = 2 D_\theta 
{\rm e}^{-\kappa t^2} \int_0^t {\rm e}^{\kappa s^2} \mathrm{d} s
\label{eq_sm16}
\end{equation}
where we have used $\langle \eta(s) \eta(s^\prime) \rangle = \delta(s-s^\prime)$. For large $t$, the asymptotic behavior of the integral in Eq.~\eqref{eq_sm16} is
\begin{equation}
\int_0^t {\rm e}^{\kappa s^2} \mathrm{d}s 
\sim \frac{{\rm e}^{\kappa t^2}}{2 \kappa t} \, ,
\label{eq_sm17}
\end{equation}
and therefore asymptotically for long times the variance is given by
\begin{equation}
\langle \delta \psi^2(t) \rangle \sim \frac{D_\theta}{\kappa t} \, .
\label{eq_sm18}
\end{equation}
From this result we can conclude that the relative angle fluctuations, $\delta \psi^2(t)$, vanish asymptotically as $1/t$. 

Now we can estimate the long-time limit of the rotational MSD, $\langle \Delta \theta^2_{\textrm{plat}} \rangle = \lim_{t\to \infty} \langle \Delta \theta^2(t) \rangle$, where $\langle \Delta \theta^2(t) \rangle = \langle (\theta(t)-\theta(0))^2 \rangle$.  We can write $\theta(t) - \theta(0) = \phi(t) - \phi(0) - (\delta \psi(t) - \delta \psi(0))$. In the long-time limit, $\phi(t)$ freezes because the motion becomes asymptotically radial, while $\delta \psi(t)$ becomes negligible.

The rotation can be also expressed as
\begin{equation}
\langle \Delta \theta^2(t) \rangle = 2 \int_0^t (t-s) C_{\dot{\theta}}(s) \, ,
\label{eq_sm19}
\end{equation}
where $C_{\dot{\theta}}(|u-v|) = \langle \dot{\theta}(u) \dot{\theta}(v) \rangle$.  Now, in our asymptotic ``locking approximation'', we replace the angular dynamics, as described by Eq.~\eqref{eq_sm3} by Eq.~\eqref{eq_sm14} for $\delta \psi(t)$.  From the solution \eqref{eq_sm15} of Eq.~\eqref{eq_sm14}, one finds $C_{\dot{\theta}}(s) = D_\theta {\rm e}^{-\kappa s^2}$ and with Eq.~\eqref{eq_sm19} the rotational MSD is $\langle \Delta \theta^2(t) \rangle = 2 D_\theta \int_0^t (t-s) {\rm e}^{-\kappa s^2} \mathrm{d}s$. In the limit $t\to \infty$ this becomes
\begin{eqnarray}
\langle \Delta \theta^2_{\textrm{plat}} \rangle & = & 
2 D_\theta \int_0^\infty {\rm e}^{-\kappa s^2} \mathrm{d}s \nonumber \\
& = & 2 D_\theta \sqrt{\frac{\pi}{\kappa}} \nonumber \\
& = & 2 \frac{k_{\textrm{B}}T}{K} 
\sqrt{\frac{\pi \gamma_{\textrm{tr}}}{\gamma_\theta}} 
\label{eq_sm20}
\end{eqnarray}
This result is confirmed by the simulation up to a numerical prefactor ($\approx 4$ in the simulation vs.~$2\sqrt{\pi}$ in our theory, cf.~the discussion of Fig.~1 in the main manuscript). This discrepancy is probably due to the approximation introduced by Eq.~\eqref{eq_sm10} where we have completely neglected the contribution due to the radial noise.

%
